\documentclass[12pt]{article}
\usepackage{amssymb}
\usepackage{bm}
\usepackage{a4}
\usepackage{graphicx}
\usepackage{comment}
\usepackage{color}

\evensidemargin \oddsidemargin
\newcommand{\RR}{\mathbb{R}}
\newcommand{\ZZ}{\mathbb{Z}}
\newcommand{\NN}{\mathbb{N}}

\def\1{{\bf 1}}
\def\0{{\bf 0}}

\def \vv {{\sf v}_2}

\def \V {{\sf V}}
\def \I {{\cal I}}

\def \LL {{\sf v}_1}

\def \T {{\cal T}}

\newcommand{\omd}{\omega_{\scriptscriptstyle d}}
\newcommand{\omu}{\omega_{\scriptscriptstyle u}}

\newcommand{\tomega}{\tilde{\omega}}

\newcommand{\homega}{\hat{\omega}}
\newcommand{\hv}{\hat{\zeta}}
\newcommand{\hI}{\hat{\I}}
\newcommand{\hpsi}{\hat{\psi}}
\newcommand{\q}{v}
\newcommand{\ti}{t}

\def\nn{\nonumber \\}

\newcommand{\be}{\begin{equation}}
\newcommand{\ee}{\end{equation}}
\newcommand{\bea}{\begin{eqnarray}}
\newcommand{\eea}{\end{eqnarray}}
\newcommand{\ba}{\begin{array}}
\newcommand{\ea}{\end{array}}
\newtheorem{prop}{Proposition}
\newtheorem{lemma}{Lemma}

\newtheorem{corollary}{Corollary}
\def\sq{\mbox{\rlap{$\sqcap$}$\sqcup$}}
\newenvironment{proof}[1]{\vspace{5pt}\noindent{\bf Proof #1}\hspace{6pt}}%
{\hfill\sq}
\newcommand{\bp}{\begin{proof}}
\newcommand{\ep}{\end{proof}\par\vspace{10pt}\noindent}
\begin{document}

\title{
The time-dependent harmonic oscillator revisited}

\author{  Gaetano Fiore$^{1,2}$ 
   \\    
$^{1}$ Dip. di Matematica e Applicazioni, Universit\`a di Napoli ``Federico II'',\\
Complesso Universitario  M. S. Angelo, Via Cintia, 80126 Napoli, Italy\\         
$^{2}$  
INFN, Sez. di Napoli, Complesso  MSA,  Via Cintia, 80126 Napoli, Italy\\
email: \ gaetano.fiore@na.infn.it}

\date{}
\maketitle

\begin{abstract} 
We point out a rather effective approach for solving the  time-dependent  harmonic oscillator $\ddot  q=-\omega^2 q$ under various regularity assumptions. 
Where $\omega(\ti)$ is $C^1$ this is reduced to  
Hamilton equation for the angle variable $\psi$ {\it alone} (the action variable $\I$ is obtained {\it by quadrature}). 
The fixed point theorem for the integral equation 
equivalent to the generic Cauchy problem for $\psi(\ti)$ 
yields a sequence $\{\psi^{(h)}\}_{h\in\NN_0}$ converging  to  $\psi$ rather fast;
if $\omega$ varies slowly or little, already $\psi^{(0)}$ approximates $\psi$ well
for rather long time lapses. 
The discontinuities of $\omega$, if any, determine those of $\psi,\I$. 
The zeros of $q,\dot q$ are investigated via
Riccati equations. Our approach
may   simplify the study of: upper and lower bounds 
on the solutions; the stability of the trivial one; 
parametric resonance when  $\omega(\ti)$ is periodic; 
the adiabatic invariance of $\I$;  asymptotic expansions in a slow time parameter $\varepsilon$;  time-dependent driven and damped parametric oscillators; etc.
\end{abstract}


\tableofcontents

\section{Introduction}  

The equation of the time-dependent harmonic oscillator 
\bea
\ddot  q(\ti)=-\omega^2(\ti) q(\ti), \qquad\omega(\ti)>0
\label{eq1}
\eea
(we abbreviate $\dot f\equiv df/d\ti$, etc.) 
has countless applications in natural sciences. In physics, it arises e.g.  in classical and quantum mechanics, optics,  electronics, electrodynamics, plasma physics, astronomy, geo- and astro-physics, cosmology (see e.g. \cite{Lew67,LewRies69,Man96,SaxGho20,HazMei13,Lik70,Lew68PR,CouSny58,Tak82,DavQin01,Lon13,Duval:2017els,AndPre19}), possibly after reduction from  more general equations.
In the equivalent  form of Hamilton equations \ $\dot q=p$, \ $\dot  p=-\omega^2 q$ \ associated to the Hamiltonian
\bea
H(q,p;\ti):=\frac 12 \left[p^2+\omega^2(\ti) q^2\right]  \label{Ham0}
\eea
it is paradigmatic for investigating general phenomena in non-autonomous Hamiltonian  
systems, such as: i) the long-time behaviour of the solutions and of the adiabatic invariants under slow or small time-dependences; ii) the characterization of the time-dependences $\omega(\ti)$ making the   trivial solution (un)stable; in particular,  iii) in the case of periodic $\omega(\ti)$ (Hill equation), the characterization of  the  $\omega(\ti)$ leading to periodic solutions or to parametric resonance; iv) the behaviour of solutions under fast or large time-dependences; v) the balance between dissipation and forcing in time-dependent driven
and damped parametric oscillators (the latter can be reduced to (\ref{eq1}) by one of the transformations of section \ref{preli}). Solving equation (\ref{eq1}) with  
 $\omega(\ti)$ as general as possible, with high degrees of accuracy and/or for long times   is paramount both for a deeper understanding
of many natural phenomena and for developing sophisticated technological applications.
For instance, the boundedness and stability of the solutions, or the knowledge
of the tiny evolution of the adiabatic invariants under slowly varying $\omega$'s  after millions or even billions of cycles
$T=2\pi/\omega$, are crucial for many phenomena and problems in electrodynamics 
(in vacuum and plasmas) applied to geo- and astro-physics \cite{Lon13}, accelerator physics \cite{CouSny58,Tak82,DavQin01}, plasma confinement
in nuclear fusion reactors \cite{HazMei13}; 
in particular, rigorous mathematical results \cite{Kul57,Kru61,Littlewood63-64,KnoPfi66,Was3,Mey73}\footnote{For a short history and more detailed list of references about the adiabatic invariant of the harmonic oscillator see e.g. the introduction of \cite{Rob16}.
For for the general theory of the adiabatic
invariants in Hamiltonian systems see e.g. \cite{LandauLifschitz,Arnold,Hen93}.} 
on the adiabatic invariance of 
$\I=H/\omega$ 
have allowed to dramatically increase the predictive power for these and other phenomena.
A number of important classical and quantum control problems 
(like the stability of atomic clocks \cite{SaxGho20}, 
the behaviour of parametric amplifiers based on electronic \cite{Howsmi70} or superconducting
devices \cite{Lik70}, or of parametrically excited oscillations in microelectromechanical systems \cite{TurEtAl98}) are ruled by linear oscillator equations reducible to  (\ref{eq1}) (as sketched in section \ref{preli}),
where the interplay 
between time-dependent driving (parametric and/or external) and/or damping 
plays a crucial role \cite{ZhaZha16}.
Moreover, known two independent solutions of (\ref{eq1}) we can find the general solution of a linear equation of the form
\bea
\dot x=Ax+a,
\qquad  A=
\left(\!\ba{cc} 0 &1 \\ -\omega^2 & 0 \ea\!\right),\quad a=\left(\!\!\ba{l} a^1\\
a^2\ea\!\!\right),
\quad x=\left(\!\!\ba{l} x^1\\
x^2\ea\!\!\right)
\label{eq1'}
\eea
with assigned $a(\ti)$, $\omega^2(\ti)$ and unknown $x(\ti)$. In \cite{Fio-impact} 
we use a family of
equations of this type to describe the evolution of the Jacobian relating
the Eulerian to the Lagrangian variables in a family of (1+1)-dimensional 
models describing the impact of a very short and intense laser pulse
into a cold diluted plasma; incidentally, this has given us the initial motivation for 
the present study. Actually, one can reduce (see section  \ref{preli}) the resolution of 
 (\ref{eq1'}a) with a {\it generic} matrix $A$ to finding two independent solutions
(\ref{eq1}) with a related $\omega^2$. In the quantum framework (\ref{eq1}) arises e.g. as the evolution equation  of the observable $q$ (a Hilbert space operator) in the Heisenberg picture (with important applications e.g. in quantum optics \cite{Man96}), but also as the time-independent
Schroedinger equation of a particle on $\RR$ in a bounded potential $U(\ti)$ and with  energy $E$, if we interpret $\ti$ as a space coordinate, $q(\ti)$ as the wave-function (which is again valued in   numbers, but complex) and set 
$\omega^2\equiv 2m(E-U)/\hbar^2$; in particular, $E>0$ and a periodic
$U(\ti)$, leading to periodic $\omega^2(\ti)$, determine  (via Bloch theorem,  an application of the Floquet one)
 the electronic bands structure of a crystal in solid state physics. Moreover, solving the Schr\"odinger equation for 
the time-dependent (and, possibly, forced) harmonic oscillator can be reduced \cite{Hus53,Sch08} to solving (\ref{eq1}).

\begin{figure}[htb!]
 \centering
 \includegraphics[height=7cm]{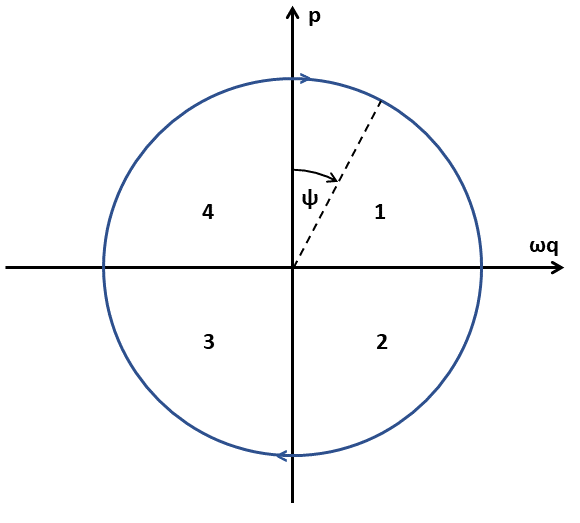}
 \caption{The angle $\psi$. We order the quadrants clockwise,
i.e. in the direction of the motion.}
 \label{anglepsi}
\end{figure}
Here we present a rather general and effective  method for approximating
the solutions of (\ref{eq1}) as accurately and as long as we wish; thereby we also derive a number of useful bounds or qualitative properties  of the solutions. We have not been  able to find this method in  the very broad  
literature (see e.g. the already cited references, the rather mathematical ones \cite{Ermakov,Pinney,Lew68JMP,
Nes13,Hat18,QinDav06}, and the references therein) on the subject.
Any advantageous analytical or numerical application to the mentioned problems and fields will contribute to its validation.
Our key observation is that, passing (section \ref{action-angle}) from the canonical coordinates $q,p$  to the angle-action coordinates $\psi,\I$ defined by
\be
\I=\frac{H}{\omega}=\frac{p^2+\omega^2q^2}{2\omega}, \qquad  \psi\, =
\, \arg\,(p,\omega q),
\label{Action-angle}
\ee
($\psi$ is thus the azimuthal angle of the generic point  in the $(p,\omega q)$ plane\footnote{Hence \ $\cot \psi=\frac {p}{\omega q}=\frac r{\omega}$, \ $\tan \psi=\frac {\omega q}{p}=s\,\omega$; \ these equations  are equivalent
where both $q,p\neq 0$, otherwise only one makes sense.}, see Fig. \ref{anglepsi})
the Hamilton equations for $\psi,\I$ 
\bea
\dot \psi =\omega +\frac{\dot\omega }{2\omega}\sin(2\psi), \qquad
\frac{\dot \I }{\I}= -\frac{\dot\omega }{\omega}\cos(2\psi) \label{eq2}
\eea
are such that the first   is {\it decoupled} from the second. This apparently overlooked feature of the harmonic oscillator [most papers do not even use the action-angle variables to analyze eq. (\ref{eq1})], among the time-dependent Hamiltonian systems admitting  angle-action variables,
is at the basis of our approach:  we reduce (\ref{eq1})  to solving  (\ref{eq2}a). In fact, afterwards (\ref{eq2}b) is  solved by quadrature, eq. (\ref{closedform-I}), and the solution of (\ref{eq1}) is obtained replacing the results in the inverse transformation of (\ref{Action-angle})
\be
 q =\sqrt{\frac{2\I}{\omega}}\,\sin\psi, \qquad p =\sqrt{2\I\omega}\,\cos\psi.
 \label{PolarCoord}
\ee
One can look for solutions of (\ref{eq2}a) numerically, or analytically, as we do here.
Reformulating the generic Cauchy problem at time $\ti_*$, (\ref{eq2}a)  with $\psi(\ti_*)=\psi_*$, 
as an integral equation and applying 
Picard iteration method 
(section \ref{approximations}),  we  obtain a sequence $\{\psi^{(h)}\}_{h\in\NN_0}$ 
such that  $|\psi\!-\!\psi^{(h)}|$ - and therefore also the corresponding 
$|\I\!-\!\I^{(h)}|$, $|q\!-\!q^{(h)}|$,
$|p\!-\!p^{(h)}|$ - uniformly go to zero in every compact interval containing $\ti_*$.
If $\omega(\ti)$ has slow or small variations, namely if \
\be
\zeta:=\dot\omega/\omega^2                                          \label{defv}
\ee
($\zeta$ is a dimensionless function measuring the relative variation of $\omega$ in the characteristic time  $1/\omega$) fulfills $|\zeta|\ll 1$, then, by (\ref{sigma_h-bound})$_{h=0}$, 
already  the 0-th order (in $\zeta$) approximation
\bea
\ba{ll}
\I^{(0)}(\ti)=\I_*:=\I(\ti_*), \qquad
&\displaystyle\psi^{(0)}(\ti)=\psi_* +\!\int^\ti_{\ti_*}\!\!\!dz\:\omega(z),\\[10pt]
\displaystyle  q^{(0)} =\sqrt{\frac{2\I_*}{\omega }}\,\sin\!\left[\psi^{(0)} \right],  \qquad 
&\displaystyle p^{(0)} =\sqrt{2\I_*\omega } \,\cos\!\left[\psi^{(0)} \right]=\dot q^{(0)} ,
\ea    \label{0th-approx}
\eea
is pretty good for quite long time intervals containing $\ti_*$, and much better than the one
\be
\tilde q(\ti):= \sqrt{\!\frac{2\I_*}{\omega_*}}\sin\!\left[\psi_* \!+\!\omega(\ti)(\ti\!-\!\ti_*)\right]
, \quad
\dot{\tilde q}(\ti)=[\omega(\ti)\!+\!\dot \omega(\ti)(\ti\!-\!\ti_*)] \sqrt{\!\frac{2\I_*}{\omega_*}}\cos\!\left[\psi_* \!+\!\omega(\ti)(\ti\!-\!\ti_*)\right]             \label{tilde-approx}
\ee
obtained from the solution of (\ref{eq1}) with constant $\omega\equiv\omega_*:=\omega(\ti_*)$ by the naive substitution $\omega_*\mapsto\omega(\ti)$. For shorter time lapses (\ref{0th-approx}) gives a reasonable approximation also if $\omega$ varies not so slowly nor little, 
see   fig.s \ref{Mathieu-nonresonant-epsilon02-a05}, \ref{Mathieu-resonant-epsilon2-a02},
 for example.
Actually, without further integrations one immediatey obtains via (\ref{def-hat-q}) a better approximation $\hat q$ of $q$, which is so to say intermediate between $q^{(0)}$ and $q^{(1)}$.
Eq. (\ref{eq2}) make sense in intervals  where $\omega $ is continuously differentiable (or is so piecewise, while keeping continuous and with bounded derivative); but all solution $\big(\psi,\I\big)$  
can be extended across the discontinuities of $\omega$  (if any)
via related matching conditions.

The plan of the paper, beside section \ref{action-angle}, is as follows. In section  \ref{preli}, while fixing the notation, we recall for which $A$, and how,  (\ref{eq1'}a)  can be reduced to a homogeneous system $\dot u=Au$
with $A$ of the form (\ref{eq1'}b), i.e. (setting $q=u^1$) to (\ref{eq1}); we also  briefly mention some possible physical applications of the reduction. 
In section  \ref{zeroes} we prove that if $
\inf_{\ti\in\RR}\{\omega(\ti)\}>0$ 
then each solution $\ti\in\RR\mapsto q(\ti)\in\RR$ admits a sequence $\{\ti_h\}_{h\in\ZZ}\subset\RR$  of interlacing  zeros of $q,\dot q$ (these play the role of $\ti_h=\frac{h\pi}{2\omega}$ in the  $\omega\!=$const  case), and we study how these $\ti_h$ depend on the initial condition fulfilled 
by $q(\ti)$; rather than via (\ref{eq2}), we do this by  reducing  (\ref{eq1})  patch by patch to a 
Riccati equation, which is well-defined also if $\omega$ is not continuous. 
In section  \ref{Useful} we sketch how our approach may help for several purposes: proving  the adiabatic invariance of $\I$ (section \ref{AdiabSect}) 
more directly; 
determining the asymptotic expansions  of $\psi,\I$ in a slow time parameter $\varepsilon$ (section \ref{AsymptSect}) faster;
determining upper/lower bounds on the solutions (section \ref{bounds});  studying the stability of the trivial one, or the occurrence 
of parametric resonance in the case of a periodic $\omega(\ti)$  (section \ref{ParametricRes}).
In section  \ref{conclu} we summarize our results, compare them to the literature, list some possible domains of application, point out open problems and directions for  further investigations.
When possible we have concentrated tedious proofs in the appendix \ref{App}.

\section{Reduction of  (\ref{eq1'}) and other equations to eq. (\ref{eq1})}
\label{preli}

Let  $\q_1,\q_2$ be the solutions of  (\ref{eq1}) fulfilling $\q_1(0)=\dot \q_2(0)=1$,
$\q_2(0)=\dot \q_1(0)=0$. The  fundamental matrix solution  of  $\dot V=AV$
 and its inverse are given by 
\be
V=\left(\!\ba{cc} \q_1 & \q_2 \\ \dot \q_1 &  \dot \q_2 \ea\!\right),       \qquad
 V^{-1}=\left(\!\ba{cc} \dot \q_2 & -\q_2 \\ -\dot \q_1 &  \q_1 \ea\!\right).     \label{matrix-sol}
\ee
In fact, from  $V(0)=I_2$  (the unit  matrix)  and $A$ having zero trace it follows  that the Wronskian $W:=\det V$  is $W=1$, namely
\be
\q_1 \dot \q_2-\q_2\dot \q_1=1.               \label{Wro}
\ee
The family $\V(t,t_*)$ (parametrized by $\ti_*$)
of $2\!\times\! 2$ matrix solutions of  
$\dot \V=A\V$ 
fulfilling $\V(\ti_*,\ti_*)=I_2$ 
is given by
\be
\V=\left(\!\ba{cc} \LL & \vv \\ \dot \LL &  \dot \vv \ea\!\right), 
 \label{DecoGG}
\ee
where $\LL,\vv$ are the following combinations of $\q_1,\q_2$: 
\bea
\ba{ll}
\vv(\ti,\ti_*)=\q_2(\ti)  \q_1(\ti_*)-\q_1(\ti)\q_2(\ti_*), \\[6pt]
\LL(\ti,\ti_*)=\q_1(\ti) \dot \q_2(\ti_*)- \q_2(\ti)  \dot \q_1(\ti_*).
\ea \label{Deco}
\eea
These fulfill also \  $ \partial \LL/\partial \ti_*|_{\ti=\ti_*}=0$, \ $\partial \vv/\partial \ti_*|_{\ti=\ti_*}=-1$, \ and  \ $\partial^2\vv/\partial \ti_*^2 =-\omega^2(\ti_*)\vv$
(whereas $\partial^2\LL/\partial \ti_*^2$ 
is more complicated). One can express $\q_2$ in terms of  $\q_1$ (and viceversa)  solving (\ref{Wro}), but the expression becomes
more and more complicated as $\ti$ gets far from 0\footnote{Known $\q_1(\ti)$ we can express 
$\q_2(\ti)$ in $]0,\ti_2[$ and $[\ti_2,\ti_3[$ respectively as  the central or right expression in
\bea
\q_2(\ti) \:\: =\:\: \q_1(\ti)\int^\ti_0\!\!\frac{dz}{\q_1^2(z)} \:\: =\:\:  \q_1(\ti)\left[\int^{\bar {\ti}}_0\!\!\frac{dz}{\q_1^2(z)}+\int^\ti_{\bar {\ti}}\!\!dz\, \frac{\omega^2(z)}{\dot \q_1^2(z)}+\frac{1}{\dot \q_1(\bar \ti)\q_1(\bar \ti)}\right]-\frac{1}{\dot \q_1(\ti)}     \label{inter}
\eea
with some  $\bar \ti\in]0,\ti_2[$; $\{\ti_h\}_{h\in\ZZ}$, with $\ti_1=0$, is the sequence of 
zeros 
of $\q_1,\dot \q_1$ studied by Proposition \ref{SpecialPoints}. The proof of the first equality is straightforward. 
The divergence of the integrand (and of the integral) as $\ti\to \ti_2^-$ is compensated by the vanishing of 
$\q_1(\ti)$, in agreement with  $\q_2(\ti_1)=-1/\dot \q_1(\ti_1)$, which follows from (\ref{Wro}).
This is manifest noting that  it is $0<\q_1(\bar \ti)<\infty$, and, using (\ref{inter}a) and integrating by parts,
\bea
\frac{\q_2(\ti)}{\q_1(\ti)}-\int\limits^{\bar {\ti}}_0\!\!\frac{dz}{\q_1^2(z)}
=\int\limits^\ti_{\bar {\ti}}\!\!\frac{dz}{\q_1^2(z)}
=-\int\limits^\ti_{\bar {\ti}}\!\!\frac{dz}{\dot \q_1(z)}\frac d{dz}\frac{1}{\q_1(z)}=\left.\frac{-1}{\dot \q_1(z)\q_1(z)}\right|_{\bar {\ti}}^y
-\int\limits^\ti_{\bar {\ti}}\!\!\frac{dz}{\q_1(z)}\frac{\ddot \q_1(z)}{\dot \q_1^2(z)}
=\left.\frac{1}{\dot \q_1(z)\q_1(z)}\right|^{\bar {\ti}}_{\ti}+\int\limits^\ti_{\bar {\ti}}\!\!dz\frac{\omega^2(z)}{\dot \q_1^2(z)}
\nonumber
\eea
(the last integral is manifestly finite as $\ti\to \ti_2$), whence  the second equality in (\ref{inter}).
The right expression and its derivative are actually well-defined for all $\ti\in]\bar {\ti},\ti_3[$.
Arguing in a similar manner one can express $\q_2(\ti)$ in other intervals $]\ti_h,\ti_{h+2}[$ as a sum
of a number of terms increasing with $|h|$.
}; 
it is more convenient to look for both $\q_1,\q_2$ globally in terms of the  angle-action variables.

For generic $a,A$, the  solution of the Cauchy problem  (\ref{eq1'}) 
with $x(\ti_*)\!=\!x_*$ can be expressed as  
\be
x(\ti)=\V(\ti,\ti_*)\left[x_*+\int^\ti_{\ti_*}\!dz\:\V^{-1}(z,\ti_*) a(z)\right],        \label{GenLin}
\ee
where $\V(\ti,\ti_*)$ has been defined in (\ref{DecoGG}).
Actually,  given a non-diagonal $\tilde A$ 
one can transform  $\dot{\tilde x}=\tilde A\tilde x\!+\!\tilde a$  into (\ref{eq1'}),
with $A$ of the form (\ref{eq1'}b), but $\omega^2$ not necessarily positive;
under additional assumptions on $A$ it is $\omega^2>0$, so that one can apply also the resolution formula (\ref{GenLin}). In fact, if $\tilde A^1_2\neq 0$  let $\Lambda$ be a solution  of the equation \
$2\dot\Lambda\!+\!\tilde A^1_1\!+\!\tilde A^2_2\!+\!\dot {\tilde A^1_2}/\tilde A^1_2=0$; \  the Ansatz
$$
x :=e^\Lambda B \tilde x, \qquad a :=e^\Lambda B \tilde a, \qquad B:=
\left(\!\ba{cc} 1 &0 \\ b & \tilde A^1_2 \ea\!\right), \quad b:=\dot\Lambda\!+\!\tilde A^1_1=\frac 12\left(\tilde A^1_1
\!-\!\tilde A^2_2\!-\! \frac {\dot {\tilde A^1_2}}{\tilde A^1_2}\right)
$$
does the job, with \
$-\omega^2
:=\dot b+b^2+ \tilde A^1_2\tilde A^2_1$. \ This must be negative in order that we may apply our resolution procedure, e.g.  $-\tilde A^1_2\tilde A^2_1>0$ must be  sufficiently large. 

As an application, consider the equations of motion \ $m\dot q= \pi$, \ $\dot \pi=-\kappa q-\eta \dot q+f$ of a particle with mass $m$  subject to an elastic force $-\kappa q$, a viscous one $-\eta \dot q$ and
 a forcing one $f$, with time-dependent $m,\kappa>0$, $\eta\ge 0$, $f$; this can be put in the form $\dot{\tilde x}=\tilde A\tilde x\!+\!\tilde a$ setting 
\be
\tilde x :=
\left(\!\ba{c} q \\ \pi \ea\!\right), \quad \tilde a :=
\left(\!\ba{c} 0 \\ f \ea\!\right), \quad\tilde A=\left(\!\ba{cc} 0 &\frac 1m \\ -\kappa &- \frac {\eta}m \ea\!\right).
\label{damped-forced}
\ee
We find \
$\dot \Lambda =b=\eta/2m$, 
\ $\omega^2=\kappa/m\!-\!\dot \eta/2m\!-\!\eta^2/4m^2$, \ which is positive provided  $\kappa>\dot \eta/2\!-\!\eta^2/4m$. Then we can reduce (\ref{damped-forced}) to (\ref{eq1'}a). Another application may be to the system
\bea
&& x''(s)+ \left[\frac 1{\rho^2(s)}-k(s)\right]x(s)=\frac{\Delta p}{\rho(s) p_0},  \label{EqTransvDyn1}\\[4pt]
&&  y''(s)+ k(s) y(s)=0,   \label{EqTransvDyn2}
\eea
which rules the transverse  beam dynamics in particle accelerators (see e.g.  \cite{Wie}).
Here: the independent parameter $s$ (replacing $\ti$) is the curvilinear abscissa along the wished particle trajectory  (design orbit) $\gamma$; $x,y$ together with $s$ make up an orthogonal set of curvilinear 
coordinates
such that $\gamma$ is characterized by the equations $x=y=0$;
$\rho(s),k(s)$ encode the first magnetic field multipoles ($\rho$ determines the curvature radius of $\gamma$);
$\Delta p$ is the momentum deviation from the reference momentum $ p_0$. The equations are valid 
in the approximation of: paraxial optics ($x,y\ll\rho$, $\ddot s \simeq 0$); 
linear beam optics (only linear terms in $x,y$ are kept in the equations);   linear field changes (represented by the first two magnetic multipoles 
$\rho(s),k(s)$);  flat accelerators ($\gamma$ lies in a plane);
quasi monochromatic beam ($\Delta p \ll p_0$).
Eq. (\ref{EqTransvDyn2}) is already of the type  (\ref{eq1}), while  eq. (\ref{EqTransvDyn1}) can be treated as in
the previous application, see (\ref{damped-forced}).

If $\tilde A^2_1\neq 0$ one can do the transformation with the indices $1,2$ exchanged.
One can reduce $\tilde A$ to the form (\ref{eq1'}b) also by a change of the `time' variable.

\section{Reducing (\ref{eq1}) to Riccati equation, and zeros of $q,\dot q$}
\label{zeroes}

The first reduction of  (\ref{eq1}) to a single  first order ordinary differential equation 
(ODE) is based on the well-known
relation between linear homogeneous second order ODE's and Riccati equations, and makes sense also 
if $\omega(\ti)$ is not continuous. We use it to study the zeros of $q$, $p=\dot q$.
Given a solution $q(\ti)$ of (\ref{eq1}), we  define $r$ and its inverse $s$ by
\be
r=\frac {\dot q}q,\qquad\qquad	 s= \frac q{\dot q} 
\label{defrs}
\ee
respectively  in an interval $K$ where $q$,  $\dot q $ does not vanish.
There $r$ (resp. $s$) fulfills the Riccati  equation
\be
\dot r=-\omega^2-r^2, \qquad\qquad	  \mbox{( resp. }\:\: \dot s=1+\omega^2\,s^2  \:\: \mbox{ )}
\label{rseqs}
\ee
 and is strictly decreasing  (resp.  growing).
In each such interval (\ref{eq1'})  is equivalent to the first order system
(\ref{defrs}a-\ref{rseqs}a) in the unknowns $r,q$
[resp. (\ref{defrs}b-\ref{rseqs}b) in the unknowns $s,q$].
Eq. (\ref{rseqs}) has the {\it only} unknown $r(\ti)$ 
[resp. $s(\ti)$]. 
Once it is solved, solving  (\ref{defrs}) for $q(\ti)$ we obtain
\bea 
q(\ti)=q_*\, \exp\left\{\int^\ti_{\ti_*}\!\!\!dz\:r(z)\right\},\qquad\qquad
\left( \mbox{resp. }\:\: 
q(\ti)=q_*\, \exp\left\{\int^\ti_{\ti_*}\!\! \frac {dz}{s(z)}\right\}  \:\:  \right).      \label{closedform-q}
\eea
Here we have parametrized the solution through its values at some point $\ti_*\in K$:  $q_*:=q(\ti_*)$ and $r_*:=r(\ti_*)=q_*/p_*$
 or  $s_*:=s(\ti_*)=p_*/q_*$ respectively, where $p_*:=p(\ti_*)=\dot q(\ti_*)$.
Finally, $\dot q(\ti)$ is obtained replacing these results again in (\ref{defrs}).
Summing up, locally we can reduce the resolution of  (\ref{eq1'}) 
to that of a {\it single}  first order equation [(\ref{rseqs}a)  or (\ref{rseqs}b)] of Riccati type. The associated Cauchy problem is equivalent to the Volterra-type integral
equation
\be
r(\ti)=r_*-\!\int^{\ti}_{\ti_*}\!\!\!dz \left[r^2\!+\!\omega^2\right]\!(z), \qquad\qquad	  \left(\mbox{resp. }\:\: s(\ti)=s_*+\!\int^{\ti}_{\ti_*}\!\!\!dz \left[1\!+\!\omega^2s^2\right]\!(z)  \:\:  \right).
\label{rsinteqs}
\ee

\medskip
The zeros of $q,p$  interlace. More precisely, in the appendix we prove

\begin{prop}
If \ $\omega_l:=
\inf_{\ti\in\RR}\{\omega(\ti)\}>0$, 
then every nontrivial 
solution $\ti\in\RR\mapsto  q(\ti)$ of (\ref{eq1}) admits a strictly increasing sequence 
$\{\ti_h\}_{h\in\ZZ}\subset\RR$ such that for all $j\in\ZZ$:
\begin{enumerate}

\item $q$ vanishes,  and $p$ has a positive maximum at $\ti=\ti_{4j}$; 

\item  $p$ vanishes,  and $q$ has a positive maximum at $\ti=\ti_{4j+1}$;

\item $q$ vanishes, and $p$ has a  negative minimum at $\ti=\ti_{4j+2}$; 

\item $p$ vanishes,  and $q$ has a  negative minimum  at $\ti=\ti_{4j+3}$;

\item $\big( q(\ti),p(\ti)\big)$  belongs to the first, second, third, fourth quadrant of the $(q,p)$ phase plane  for all $\ti$  respectively in $]\ti_{4j},\ti_{4j+1}[$, $]\ti_{4j+1},\ti_{4j+2}[$,  $]\ti_{4j+2},\ti_{4j+3}[$,  $]\ti_{4j+3},\ti_{4j+4}[$.
\end{enumerate}
More generally, if $\omega(\ti)\ge\bar\omega_l$, with some $\bar\omega_l>0$,  holds for all 
$\ti$ belonging to an interval $J\subseteq \RR$,
 then there is a subset of consecutive integers $\ZZ_J\subseteq\ZZ$ such that 
1. to 5.
 hold for all $\ti_h$, $h\in\ZZ_J$. 
\label{SpecialPoints}
\end{prop}

\noindent 
This generalizes the case  $\omega\!=$const, with
$ q(\ti)=\sin(\omega \ti)$, $p(\ti)=\omega\cos(\omega \ti)$, and $\ti_h=\frac{h\pi}{2\omega}$. 
The labelling of these special points is defined up to a shift $h\mapsto h+4k$, with a fixed $k\in \ZZ$.

Under the assumptions of proposition \ref{SpecialPoints} the function $r$ 
is defined and strictly decreasing in each interval \  $]\ti_{2k},\ti_{2k+2}[$,  \ 
diverges at the extremes and vanishes at the middle point $\ti_{2k+1}$;
in each such interval (\ref{eq1})  is equivalent to the first order system
(\ref{defrs}a-\ref{rseqs}a) in the unknowns $r,q$.
Similarly, $s$ is  defined and strictly growing  in each interval  $]\ti_{2k-1},\ti_{2k+1}[$
diverges at the extremes, and vanishes at the middle point  $\ti_{2k}$; 
in each such interval (\ref{eq1})  is equivalent to the first order system
(\ref{defrs}b-\ref{rseqs}b) in the unknowns $s,q$.

If $\omu,\omd$ are positive constants such that $\omd\le\omega(\ti)\le\omu$ for 
$\ti\in]\ti_h,\ti_{h+1}[$, then we easily obtain the following rough bounds on the length of this interval\footnote{In fact, by these bounds  every solution $r$ of (\ref{rseqs})  in a 
neighbourhood of any $\ti_*\!\in]\ti_{2k},\ti_{2k+2}[$ fulfills 
\bea
-\dot r\ge \omd^2+r^2 \quad\Rightarrow\quad	
\frac{-\dot r/\omd}{1+(r/\omd)^2}=\frac d{dy} \cot^{-1}\!\left(\!\frac r{\omd}\!\right)\ge \omd \quad\Rightarrow\quad	\cot^{-1}\!\left[\frac {r(\ti)}{\omd}\right]- \cot^{-1}\!\left[\frac {r(\ti_*)}{\omd}\right]\ge \omd (\ti\!-\!\ti_*) \nn
-\dot r\le \omu^2+r^2 \quad\Rightarrow\quad	
\frac{-\dot r/\omu}{1+(r/\omu)^2}=\frac d{dy} \cot^{-1}\!\left(\!\frac r{\omu}\!\right)\le \omu \quad\Rightarrow\quad	\cot^{-1}\!\left[\frac {r(\ti)}{\omu}\right]- \cot^{-1}\!\left[\frac {r(\ti_*)}{\omu}\right]\le \omu (\ti\!-\!\ti_*).
\nonumber
\eea
Choosing $\ti=\ti_{2k+2}$ and   $\ti_*=\ti_{2k+1}$ 
 leads to (\ref{Delta\ti_hbound}) 
for $h=2k+1$; choosing $\ti=\ti_{2k+1}$ and   $\ti_*=\ti_{2k}^+$  leads to (\ref{Delta\ti_hbound}) for $h=2k$.
Here we have used $r(\ti_{2k+1})=0$, $r(\ti_{2k+2}^-)=-\infty$, $r(\ti_{2k})=+\infty$.
}:
\be
\frac {\pi}{2\omu} \le \ti_{h+1}-\ti_h \le \frac {\pi}{2\omd}.        \label{Delta\ti_hbound}
\ee
These inequalities are most stringent   if we adopt 
$$
\omu\equiv \sup_{]\ti_h,\ti_{h+1}[}\left\{\omega(\ti)\right\}, \qquad
\omd\equiv \inf_{]\ti_h,\ti_{h+1}[}\left\{\omega(\ti)\right\}.
$$
In general we are not able to determine such
$\sup,\inf$, because we don't know the exact locations of $\ti_h,\ti_{h+1}$. However the choice of
$\omu,\omd$ 
can be improved recursively\footnote{Assuming 
for simplicity that $\ti_h$ is known, 
a candidate $\omu$ can be accepted  if $\omu\ge\sup_{]\ti_h,\ti_h\!+\!\frac{\pi}{2\omu}[}\left\{\omega(\ti)\right\}$, must be rejected otherwise; if the inequality is strict a better candidate
will be $\omu'\equiv\sup_{]\ti_h,\ti_h\!+\!\frac{\pi}{2\omu}[}\left\{\omega(\ti)\right\}$; and so on.
Similarly one argues for $\omd$.}.
More stringent bounds on the length of the interval will be determined in section \ref{bounds}.

\bigskip
Now let $Q(\ti;\ti_*,q_*,p_*)$ be the family of solutions of (\ref{eq1}) fulfilling the  conditions
\be
Q(\ti_*;\ti_*,q_*,p_*)=q_*,\qquad \dot Q(\ti_*;\ti_*,q_*,p_*)=p_*.
\label{InCondGen}
\ee
We ask how the sequence  of special points  $\{\ti_h\}_{h\in \ZZ}$ for $Q$ (in the sense of Proposition \ref{SpecialPoints})  depends on the parameters $(\ti_*,q_*,p_*)\in\RR^3$. 
Here we partly investigate this question.
First we note that, since $Q(\ti_*;\ti_*,aq_*,ap_*)=a\,Q(\ti_*;\ti_*,q_*,p_*)$, then 
\  $s:= Q/\dot Q$,   $r:=\dot Q/ Q$ and therefore also the sequence is
invariant under all rescalings $\big(q_*,p_*\big)\mapsto \big(a\,q_*,\, a\, p_*\big)$, $a\in\RR^+$;
the latter map each quadrant of the $(q,p)$ plane into itself. 
In the appendix we prove

\begin{prop} \ For all $(q_i,p_i)\in\RR^2\setminus \{(0,0)\}$, all the 
special points $\ti_h$ associated to the
family of solutions $Q(\ti;\ti_i,q_i,p_i)$, seen as  functions of $\ti_i$, are strictly growing.
\label{MonotonSpecialPoints}
\end{prop}

This applies in particular to the $\{\ti_h(\ti_i)\}_{h\in \ZZ}$ of the families of solutions 
$Q=\LL,\vv$ defined by  (\ref{Deco}); we can remove the residual ambiguity  $h\mapsto h+4k$ in their definitions
setting $\ti_0(\ti_i)\!=\!\ti_i$ for $\vv(\ti,\ti_i)$, and $\ti_1(\ti_i)\!=\!\ti_i$
for $\LL(\ti,\ti_i)$.   If  $\omega\!=$const it is \ $\vv(\ti,\ti_i)=\sin[\omega(\ti\!-\!\ti_i)]$, \
$\ti_h(\ti_i)=\ti_i+\frac{h\pi}{2\omega}$\ and \ $\LL(\ti,\ti_i)=\cos[\omega(\ti\!-\!\ti_i)]$, \
$\ti_h(\ti_i)=\ti_i+\frac{(h\!-\!1)\pi}{2\omega}$ \ respectively.

\bigskip
Chosen any compact interval $K_k\!\subset\,]\ti_{k},\ti_{k+2}[$, applying the fixed point theorem one can recursively build via
\be
r_k^{(h)}(\ti):=
r_k^*-\!\int^{\ti}_{\ti_k}\!\!\!dz \left[\left(r^{(h-1)}\right)^2\!+\!\omega^2\right]\!(z), \qquad   \left(\mbox{resp. }\:\: s_k^{(h)}(\ti):=s_k^*+\!\int^{\ti}_{\ti_*}\!\!\!dz \left[1\!+\!\omega^2\left(s^{(h-1)}\right)^2\right]\!(z)     \right)
\label{rs-h}
\ee
a sequence $\{r_k^{(h)}\}_{h\in\NN_0}\subset C(K)$ if $k$ is even (resp. a sequence $\{s_k^{(h)}\}_{h\in\NN_0}\subset C(K)$  if $k$ is odd) that
converges uniformly  to the unique solution $r_k$ (resp. $s_k$) of (\ref{rsinteqs}) in $K_k$\footnote{One possibility is to start with $r_k^{(0)}:=r_k^*$  (resp. $s_k^{(0)}:=s_k^*$);
this is especially convenient  when $\omega$ is 'small'. The corresponding 
first order approximations are
\bea
q_k^{(0)}(\ti)=q_k^*\, \exp\left\{r_k^*(\ti\!-\!\ti_*)\right\},
\qquad r_k^{(1)}(\ti)= 
r_k^* - r_k^*{}^2(\ti\!-\!\ti_*)-\!\int^{\ti}_{\ti_*}\!\!\!dz\, \omega^2(z), \nn
q_k^{(1)}(\ti)=q_k^*\, \exp\left\{r_k^*(\ti\!-\!\ti_*)- \frac{r_k^*{}^2}2(\ti\!-\!\ti_*)^2-\!\int^{\ti}_{\ti_*}\!\!\!dz\, (t\!-\!z)\,\omega^2(z)\right\},\\
\left(\mbox{resp. }\:\: q_k^{(0)}(\ti)=q_k^*\, \exp\left\{\frac{\ti\!-\!\ti_*}{s_k^*}\right\},
 \qquad   s_k^{(1)}(\ti)=s_k^*+(\ti\!-\!\ti_*)+ s_k^*{}^2\!\int^{\ti}_{\ti_*}\!\!\!dz \,\omega^2(z)  \quad   \right.,\nn 
\left. q^{(1)}(\ti)=q_k^*\,  \exp\left\{\int^\ti_{\ti_*}\!\! \frac {dz}{s^{(1)}(z)}\right\}
\quad  \right).
\eea
In particular, we find that if the initial conditions are $q(0)=0$, $\dot q(0)=1$ then  $q(\ti)\simeq \ti$ for small $t$.
If the initial conditions are $q(0)=1$, $\dot q(0)=0$ then  $q(\ti)\simeq q^{(1)}(\ti)=\exp\left\{-\!\int^{\ti}_{0}\! dz\, (t\!-\!z)\,\omega^2(z)\right\}$ for small $t$.
}.
Replacing $r$ (resp. $s$) by $\{r_k^{(h)}\}$  (resp. $\{s_k^{(h)}\}$) 
  in (\ref{closedform-q}) we obtain the associated sequence
$\{q_k^{(h)}\}_{h\in\NN_0}\subset C(K)$. Incidentally, we note that such a resolution procedure
may be applied also if $\omega$ vanishes at some $\ti\in K$. To obtain 
a solution of (\ref{eq1}) in all of $\RR$ in this way one can choose all two consecutive
intervals $K_k$ with a non-empty intersection and match the initial values $r_k^*$ (resp. $s_k^*$) in $K_k$ so that the globally defined $q,\dot q$ are continuous. On the contrary,
solving the first order ODE (\ref{eq2}) yields the global solution
$q(t)$ of  (\ref{eq1}) at once.  This is discussed in the next sections.

\section{Solving Hamilton equations for action-angle variables}
\label{action-angle}

\subsection{Reduction to the Hamilton equation for  the angle variable}
\label{reduction}

To start, the reader may check by a direct computation that  (\ref{eq1}) and  (\ref{eq2})
are equivalent.  The root of this equivalence is the fact that the  transformation $(q,p)\mapsto (\I,\psi)$ to the  action, angle variables defined by (\ref{Action-angle}) is canonical 
even  if $\omega$ - and therefore also $H$ and the transformation itself - depend on time\footnote{The calculation
leading to the Poisson bracket $\{\psi,\I\}\!=\!1$ holds regardless of the time dependence of 
$\omega$.}.   Eq. (\ref{eq2}) are nothing but the Hamilton equations satisfied by $\I,\psi$
\bea
\dot \psi =\frac{\partial K}{\partial \I}, \qquad
\dot \I= -\frac{\partial K}{\partial \psi}, \qquad\quad K=\left[\omega+\frac{\dot\omega }{2\omega}\sin(2\psi)\right]\I; \nonumber
\eea
the transformation $(q,p,H)\mapsto (\I,\!\psi,\!K)$, including 
the transformed Hamiltonian $K(\I,\psi,\ti)$,
are obtained from the generating function $F(q,\psi,\ti)$ (of type 1) via
\bea
I=-\frac{\partial F}{\partial \psi},\qquad p=\frac{\partial F}{\partial q}, \qquad 
K:=H+\frac{\partial F}{\partial t}.    \label{GenFun}
\eea
$F$ must  fulfill \ $p\,dq-H dt=\I\, d\psi-Kdt+dF$, \ see e.g. \cite{LandauLifschitz}.  The latter equation is solved  by \ $F(q,\psi,\ti)=\frac {\omega(\ti)}2\, q^2\cot\psi$. \
Replacing the latter in (\ref{GenFun}) we indeed obtain $K$ and (\ref{Action-angle}).

From (\ref{PolarCoord}) it  follows that the sequence  of special points  $\{\ti_h\}_{h\in \ZZ}$  (in the sense of Proposition \ref{SpecialPoints}) of every solution $q(\ti)$ is characterized,  up to a multiple of $2\pi$, by
\be
\psi(\ti_{h})=h \frac{\pi}2, \qquad h\in\ZZ.       \label{psi-specialvalues}
\ee
The Cauchy problem (\ref{eq2}a) with  $\psi(\ti _*)=\psi _*$ is equivalent to the Volterra-type 
integral equation
\bea
\ba{l}
\displaystyle \psi(\ti)= \varphi(\ti)+\int^\ti_{\ti_*}\!\!dz\:\left[ \frac{\dot\omega }{2\omega}\sin\left(2\psi\right)\right]\!(z),            \\[14pt]
\displaystyle \mbox{where }
\varphi(\ti):= \psi _*+\!\int^\ti_{\ti_*}\!\!dz\:\omega(z)=:\psi^{(0)}(\ti).
\ea \label{Inteq}
\eea
Once  this is solved, (\ref{eq2}b) with initial condition $\I(\ti_*)=\I_*$ is solved  by 
\be
\I(\ti)=\I_*
\exp\left\{-\!\!\int^\ti_{\ti_*}\!\!dz \left[\frac{\dot\omega }{\omega}\cos(2\psi)\right]\!(z)\right\}; \label{closedform-I}
\ee
the time-dependence of $H,q,p$ is obtained
expressing them  in terms of $\psi,\I,\omega$. 
If $|\zeta|\equiv|\dot\omega/\omega^2|\ll 1$ (slowly/slightly varying $\omega$) then the integral
in (\ref{Inteq}) can be neglected, the variation  
of $\I$ over time intervals $[\ti_*,\ti]$ containing many cycles\footnote{We define $[\ti_1,\ti_2]$ as a {\it cycle} if $\psi(\ti_2)-\psi(\ti_1)=2\pi$.} can be 
approximated replacing  $\cos(2\psi)$ by its mean
 $0$ over a cycle, whence  [cf. (\ref{0th-approx})] 
\be
\psi(\ti)\simeq \varphi(\ti),\qquad 
\I(\ti)\simeq \I(\ti_*),\qquad H(\ti) \simeq H(\ti_*)\: \omega(\ti)/\omega(\ti_*) .
\ee
The same applies if $\dot\omega/\omega$ oscillates  about zero much faster than $\varphi$; 
such oscillations almost wash out the integrals in (\ref{Inteq}), (\ref{closedform-I}).
Improved estimates of such integrals  lead to the better 
approximations reported below.

\bigskip
Eq. (\ref{eq2}) make sense in intervals  where $\omega $ is continuously differentiable, or is so at least piecewise while keeping  continuous and with a bounded derivative.
If  $\omega(\ti)$ is continuous in $]\ti_1,\ti_2[$ except at some point $\ti_d$, then in general
so are $\ddot q(\ti),\psi(\ti), \I(\ti),H(\ti)$,
whereas $q(\ti),p(\ti), r(\ti),s(\ti)$ are continuous in the whole $]\ti_1,\ti_2[$. Let
\bea
\omega_\pm \!:=\! \lim\limits_{\ti\to \ti_d^\pm}\omega(\ti),
\qquad \psi_\pm \!:=\! \lim\limits_{\ti\to \ti_d^\pm}\psi(\ti)
,\qquad\I_\pm \!:=\! \lim\limits_{\ti\to \ti_d^\pm}\I(\ti);
\eea
note that by their definitions [cf.  (\ref{Action-angle})] both $\psi_+,\psi_-$ must belong to the same quadrant of the $(q,p)$ plane.
The continuities of $r,q,p$ at the discontinuity point $\ti_d$ of $\omega(\ti)$ imply the matching relations 
\bea
\frac{\tan\psi_+}{\omega_+}=\frac{\tan\psi_-}{\omega_-},
\qquad \I_+=\I_-\,\frac{\omega_-}{\omega_+}\,\frac{1+\left(\frac{\omega_+}{\omega_-}\tan\psi_-\right)^2}{1+\tan^2\psi_-}.
\label{matched}
\eea
The second determines the `sudden' change of $\I$
due to the  `sudden' change of $\omega$; this depends not only on $\omega_\pm$,
but also on the value of $\psi_-$ (or, equivalently, $\psi_+$).
If \ $\ti_d=\ti_h$, $h\in\ZZ$, then $\psi$ remains continuous
because \ $\tan\psi(\ti_h)\!=\!0,\pm\infty$, \ while 
$$
\I_+\omega_+=\I_-\omega_-\quad\mbox{if  $h$ even}, \qquad \quad
\I_+\omega_-=\I_-\omega_+ \quad\mbox{if   $h$ odd.}
$$
Therefore the maximum, minimum possible ratio $\I_+/\I_-$ 
are resp. $\max\Omega$, $\min\Omega$, where $\Omega:=\{\omega_+/\omega_-,\omega_-/\omega_+\}$.
In general $\psi(\ti)$ can be extended from \ $]\ti_1,\ti_d[$ \ (resp.
\ $]\ti_d,\ti_2[$) \ to the whole  $]\ti_1,\ti_2[$ solving  (\ref{eq2}a) with  (\ref{matched}a)
as initial (resp. final)  conditions  in the complementary part. 
$\I(\ti)$ can be extended from \ $]\ti_1,\ti_d[$ \ (resp.
\ $]\ti_d,\ti_2[$) \ to the whole  $]\ti_1,\ti_2[$ applying  (\ref{closedform-I}) with  (\ref{matched}b)
as initial (resp. final)  conditions  in the complementary part. 

\subsection{Sequence converging to the solution of (\ref{Inteq})}
\label{approximations}

We can reformulate  (\ref{Inteq})  in every compact time interval \
$K:=[a,b]\supset \ti_*$ \  as the fixed point equation \ $\T \psi=\psi$, \
where $\T$ is the linear map $\T: C(K{})\to C(K{})$ defined by
\be
[\T u](\ti):=\displaystyle \varphi(\ti)+\int^\ti_{\ti_*}\!\!dz\:f[z,u(z)], \qquad f(\ti,u):= \frac{\dot\omega(\ti) }{2\omega(\ti)}\sin\left(2 u\right).
\ee
By means of 
Picard iteration method  (an application of Banach fixed point theorem)
 one can build
sequences $\{\psi^{(h)}\}_{h\in\NN_0}\subset C(K)$  converging to its unique solution $\psi$.
A very convenient one starts with $\psi^{(0)}:=\varphi$ and continues with
\be
\psi^{(h)}(\ti):=[\T^h\varphi](\ti)
= \varphi(\ti)+\int^\ti_{\ti_*}\!\!dz\:\left[ \frac{\dot\omega }{2\omega}\sin\left(2\psi^{(h-1)}\right)\right]\!(z).             \label{psi-h}
\ee
The corresponding sequence $\{\I^{(h)}\}_{h\in\NN_0}$ converging to $\I$
arises replacing $\psi\mapsto\psi^{(h-1)}$ in (\ref{closedform-I})
\be
\I^{(h)}(\ti) :=\I(\ti_*)
\exp\left\{-\!\!\int^\ti_{\ti_*}\!\!dz \left[\frac{\dot\omega }{\omega}\cos\left(2\psi^{(h-1)}\right)\right]\!(z)\right\},                       \label{closedform-I-h}
\ee
and the ones 
of $H,q,p$ are obtained replacing $(\I,\psi)\mapsto(\I^{(h)},\psi^{(h)})$ in (\ref{Action-angle}), (\ref{PolarCoord}).
The  `errors' of approximation are conveniently  bounded via
 the {\it total variation} of $\log\omega(\ti)$ between $\ti_*,\ti$ 
\be
g(\ti):=\int^\ti_{\ti_*}\!\!dz \left| \frac{\dot\omega (z)}{\omega(z)}\right|.
\ee
This grows with $\ti$, and $g(\ti_*)=0$. 
If  $\omega(\ti)$ is monotone in $[\ti_*,\ti]$
then $g(\ti)=|\log\frac{\omega(\ti)}{\omega(\ti_*)}|$; otherwise $g$ is  the sum of
a term of this kind for each  monotonicity interval contained in $[\ti_*,\ti]$.

\begin{prop}
If $\omega\in C^1(K)$, with $\omega>0$, then for all $h\in\NN$ 
\bea
\ba{l}
 2\left|\psi(\ti)- \psi^{(h)}(\ti)\right| \,\le\, \displaystyle  \frac{|g(\ti)|^{h+1}}{(h\!+\!1)!}\:, \\[10pt]
\left|\log\I(\ti)- \log\I^{(h)}(\ti)\right| \,\le\,  \displaystyle \frac{|g(\ti)|^{h+1}}{ (h\!+\!1)!}\: .
\ea
\label{sigma_h-bound}
\eea
Consequently, \ $\psi^{(h)}\to\psi$, \ \ $\I^{(h)}\to\I$ \   in the sup
norm  of $C(K{})$, \ $\Vert u\Vert_\infty:=\sup_{\ti\in K{}}\left\{|u(\ti)|\right\}$.
\label{main-prop}
\end{prop}

\bp{}
We start by deriving the following inequalities: for all $\alpha,\delta\in\RR$ 
\be
\ba{l}
 |\sin(\alpha\!+\!\delta)\!-\!\sin\alpha|\le \sqrt{2\left(1\!-\!\cos\delta\right)}\le |\delta|,\\[6pt]
|\cos(\alpha\!+\!\delta)\!-\!\cos\alpha|\le \sqrt{2\left(1\!-\!\cos\delta\right)}\le |\delta|.
\ea \label{LittleLemma}
\ee
We abbreviate $f(\alpha,\delta):=\sin(\alpha\!+\!\delta)\!-\!\sin\alpha=
\sin\alpha(\cos\delta\!-\!1)\!+\!\cos\alpha\,\sin\delta$. For fixed $\delta$
the points $\alpha'$ of maximum, minimum for $f$ fulfill
\bea
0=\frac{\partial f}{\partial\alpha}(\alpha',\delta)=\cos(\alpha'\!+\!\delta)\!-\!\cos\alpha'
=\cos\alpha'(\cos \delta\!-\!1)\!-\!\sin\alpha'\sin\delta\quad\Rightarrow\quad
\tan\alpha'=\frac{\cos \delta\!-\!1}{\sin\delta}\nn[6pt]
f(\alpha',\delta)=\cos\alpha'\left[(\cos\delta\!-\!1)\tan\alpha'\!+\! \sin\delta\right]
=\frac{\cos\alpha'}{\sin\delta}\left[(\cos\delta\!-\!1)^2\!+\! \sin^2\delta\right]
=2\cos\alpha'\,\frac{ 1\!-\!\cos\delta }{\sin\delta}\nn[6pt]
\Rightarrow\quad\left|f(\alpha,\delta)\right|\le \left|f(\alpha',\delta)\right|=\frac 2{\sqrt{1\!+\! \tan^2\alpha'}}\,\frac{ 1\!-\!\cos\delta }{\left|\sin\delta\right|}=\frac{2\left(1\!-\!\cos\delta\right)}{\sqrt{\sin^2\delta\!+\! \left(1\!-\!\cos\delta\right)^2}}=\sqrt{2(1\!-\!\cos\delta)},
\nonumber
\eea
as claimed; the right inequality in (\ref{LittleLemma}a) follows from the left one   and the one
\ $1\!-\!\cos\delta\le\frac{\delta^2}2$ \ (obtained integrating \
$\sin x\le x$ over $[0,\delta]$).  Inequalities (\ref{LittleLemma}b) follow via the
shift $\alpha\mapsto\alpha\!+\!\pi/2$.

By the assumptions it is $ \mu:=\max_K   \big|\frac{\dot\omega}{\omega}\big|<\infty$ and $M:=\max_K\{|g(\ti)|\}=\max\{|g(a)|,g(b)\}<\infty$. \ 
Now abbreviate \ $\sigma^{(h+1)}(\ti):=\psi(\ti)- \psi^{(h)}(\ti)$. \
Eq. (\ref{Inteq})  implies 
\bea
2\sigma^{(1)}(\ti)= \int^\ti_{\ti_*}\!\!dz
\left[ \sin (2\psi)\frac{\dot\omega}{\omega}\right]\!(z),\qquad\quad
2\left|\sigma^{(1)}(\ti)\right| \le\int^\ti_{\ti_*}\!\!dz \left| \frac{\dot\omega (z)}{\omega(z)}\right| = g(\ti)\quad \mbox{if }\: \ti\ge \ti_*,  \label{inter0}
\eea
i.e. (\ref{sigma_h-bound}) for $h=0$. If (\ref{sigma_h-bound}) holds for $h\!=\!j\!-\!1\ge 0$
 it does also for $h\!=\!j$, since by eq. (\ref{Inteq}), (\ref{psi-h})
\bea
2\sigma^{(j+1)}(\ti)  &\!=\! &
\int^\ti_{\ti_*}\!\!dz \left\{\frac{\dot\omega }{\omega}\left[\sin\left(2\psi\right) -\sin\left(2\psi^{(j-1)}\right)\right]\right\}\!(z),  \label{inter1}\\
2\left|\sigma^{(j+1)}(\ti)\right|  &\!\le\! & \int^\ti_{\ti_*}\!\!dz \left\{\left| \frac{\dot\omega }{\omega}\right| \left|\sin\left(2\psi^{(j-1)}\!+\!2\sigma^{(j)}\right) -\sin\left(2\psi^{(j-1)}\right)\right|\right\}\!(z)  \le\int^\ti_{\ti_*}\!\!dz \left|\frac{\dot\omega }{\omega} \,2\sigma^{(j)}\right|\!(z)  \nn
&\!\le\! & \int^\ti_{\ti_*}\!\!dz \,  \left[\left| \frac{\dot\omega }{\omega}\right|
\frac{g^{j}\!\!}{j!}\right]\!(z) =\int^\ti_{\ti_*}\!\!  \frac{dz}{ (j\!+\!1)!} 
\,\frac d{dz} g^{j+1} (z)=\frac{[g(\ti)]^{j+1}}{ (j\!+\!1)!}\qquad \mbox{if }\: \ti\ge \ti_*.
\nonumber
\eea
Similarly, eq. (\ref{closedform-I}), (\ref{closedform-I-h}) imply 
$\log[\I(\ti)/\I^{(j)}(\ti)]\!=\!\int^\ti_{\ti_*}\!\!dz \left\{\frac{\dot\omega }{\omega}\left[\cos\!\left(2\psi^{(j-1)}\right)\!-\!\cos\!\left(2\psi\right)\right]\right\}$, whence
\bea
\left|\log\frac{\I(\ti)}{\I^{(j)}(\ti)}\right|  &\!\le\! & \int^\ti_{\ti_*}\!\!dz \left\{\left| \frac{\dot\omega }{\omega}\right| \left|\cos\left(2\psi^{(j-1)}\right)-\cos\left(2\psi^{(j-1)}\!+\!2\sigma^{(j)}\right) \right|\right\}\!(z)  \le\int^\ti_{\ti_*}\!\!dz \left|\frac{\dot\omega }{\omega} \,2\sigma^{(j)}\right|\!(z)  \nn
&\!\le\! & \int^\ti_{\ti_*}\!\!dz \,  \left[\left| \frac{\dot\omega }{\omega}\right|
\frac{g^{j}\!\!}{(j)!}\right]\!(z) =\int^\ti_{\ti_*}\!\!  \frac{dz}{ (j\!+\!1)!} 
\,\frac d{dz} g^{j+1} (z)=\frac{[g(\ti)]^{j+1}}{ (j\!+\!1)!} \qquad \mbox{if }\: \ti\ge \ti_*,
\nonumber
\eea
as claimed. If $\ti<\ti_*$ the previous inequalities hold if we exchange the extremes of integration $\ti,\ti_*$,
thus proving (\ref{sigma_h-bound}) also in this case.   From $|g(\ti)|\le M$  we find  rhs(\ref{sigma_h-bound})$\le M^{h+1}/(h\!+\!1)!$, which goes to zero as $h\!\to\!\infty$; this proves the mentioned convergences \ $\psi^{(h)}\!\to\!\psi$,  $\I^{(h)}\!\to\!\I$. 
\ep

\begin{corollary}
If $\omega\in C^1(\RR)$, with $\omega>0$, then  \ $\psi^{(h)}\to\psi$, \ \ $\I^{(h)}\to\I$ \  
pointwise in all of $\RR$.
\end{corollary}

{\bf Remarks}: 
\begin{enumerate}

\item  The rate  of the convergence  \ $\psi^{(h)}\to\psi$ \
 for large $|\ti\!-\!\ti_*|$ is tipically much better than what the bounds (\ref{sigma_h-bound}) suggest.
This makes the approximating sequence $\{\psi^{(h)}\}_{h\in\NN_0}$  very useful also for practical purposes and numerical computations.
In fact, if $|\varphi(\ti)\!-\!\psi_*|\gg \pi$ then the integrand of  (\ref{inter0}a)  
changes sign many times in $[\ti,\ti_*]$, thus reducing the magnitude of the integral and leading to 
$2\left|\sigma^{(1)}(\ti)\right| \ll |g(\ti)|$,
 unless $\dot\omega/\omega$ oscillates  in phase with $\sin(2\varphi)$  (resonance).
In the latter case, one 
could show that  \ lhs $\ll$ rhs \  in  (\ref{sigma_h-bound})  holds at least 
for the following corrections (i.e. for $h>1$). 
This is left as a hint for future investigations.

\item Since \ $|g(\ti)|\le\mu |\ti\!-\!\ti_*|$, \ the bounds  (\ref{sigma_h-bound}) 
hold also replacing  the rhs by 
\ $[\mu |\ti\!-\!\ti_*|]^h/h!$. 

\item 
The application of the  Banach
fixed point theorem yielding the Picard-Lindel\"of theorem (see e.g. \cite{Ame86}, pp. 406-410)
also leads to \ $\Vert \psi\!-\!\psi^{(h)}\Vert_\infty\to 0$ (see the appendix), 
\ but via   
the class of pointwise bounds  (parametrized by $\lambda>\mu/2$)
\be
 \left|\psi^{(h)}(\ti)\!-\!\psi(\ti)\right| \: <\: e^{\lambda |\ti-\ti_*|}\:\frac{\nu^h}{1\!-\!\nu} \: 
 \left\| \varphi\!-\!\psi^{(1)}\right\|_{\lambda, {\scriptscriptstyle K}}
\label{sigma_h-bound-0'}
\ee
where $\nu:= \frac{\mu}{2\lambda}\in]0,1[$, $ \|u\|_{\lambda, {\scriptscriptstyle K}}:=\max_{\ti\in K{}}
\left\{|e^{-\lambda |\ti-\ti_*|}u(\ti)|\right\}$; 
the infimum $\sigma_{h,K}$ of the rhs over $\lambda\!\!\in]\mu/2,\infty[$ gives the most stringent bound in the class.
Nevertheless, (\ref{sigma_h-bound-0'})  is less  manageable and stringent  
than (\ref{sigma_h-bound}a): as $h\to\infty$ the latter goes faster to zero, also due to the factorial
in the denominator.

\item Once $\varphi$ is computed, through it we can readily use the  approximation
\be
\hat q(\ti):=
\sqrt{\frac{2\I^{(1)}(\ti)}{\omega(\ti)}}\:\sin\psi^{(0)}(\ti)=
\sqrt{\frac{2\I(\ti_*)}{\omega(\ti)}}\exp\left\{-\!\!\int^\ti_{\ti_*}\!\!dz \left[\frac{\dot\omega }{2\omega}\cos\left(2\varphi\right)\right]\!(z)\right\}\:\sin\varphi(\ti),               \label{def-hat-q}
\ee
which  
is an intermediate one between the above defined ones $q^{(0)}$ and  $q^{(1)}$.
\end{enumerate}

\noindent
In the $h$-th approximation the discontinuity relations (\ref{matched}) become
\bea
\frac{\tan\psi^{(h)}_+}{\omega_+ }=\frac{\tan\psi^{(h)}_-}{\omega_-},
\quad \frac{\I^{(h)}_+}{\omega_+}\sin^2\!\psi^{(h)}_+=\frac{\I^{(h)}_-}{\omega_-}\sin^2\!\psi^{(h)}_-,\quad
\I_+\omega_+\cos^2\!\psi^{(h)}_+=\I_-\omega_-\cos^2\!\psi^{(h)}_-.
    \label{matched-h}
\eea
Consequently, $\psi^{(h)}(\ti)$ is given by (\ref{psi-h}) for  $\ti\in]\ti_*,\ti_d[$, while for $\ti\in ]\ti_d,\ti_f[$ it is given by
 \be
\psi^{(h)}(\ti)= \psi^{(h)}_+ +\int^\ti_{\ti_d}\!\!dz\:\left[ \omega+\frac{\dot\omega }{2\omega}\sin\left(2\psi^{h-1}\right)\right]\!(z).             \label{psi-h+}
\ee
From these formulae one can  derive  an upper bound for the `error' $\sigma^{(h+1)}_+\!:=\!\psi_+- \psi^{(h)}_+$ from one for the `error' 
$\sigma^{(h+1)}_-\!:=\!\psi_-- \psi^{(h)}_-$, and conversely.

\section{Usefulness of our approach}
\label{Useful}

\subsection{Adiabatic invariance of $\I$}
\label{AdiabSect}

The action variable 
$\I=H/\omega$ of the harmonic oscillator is the simplest example of an {\it adiabatic invariant} in a Hamiltonian
system\footnote{It allowed 
among other things to interpret the Planck's quantization rule as
the first instance of the Bohr-Sommerfeld-Ehrenfest quantization rules in the socalled {\it Old Quantum Mechanics}.
At the Solvay Congress of 1911 on the old quantum theory Lorentz asked how the
amplitude of a simple pendulum would vary if its
period were slowly changed by shortening its string.
Would the number of quanta of its motion change?
Einstein answered that the action variable
$E/\omega$, where $E$ is its energy and $\omega$ its frequency, would
remain constant and thus the number of quanta would
remain unchanged, if $\dot\omega/\omega_0$ were small enough \cite{Einstein1911}.
}, in the sense that it remains approximately constant for  slowly varying frequencies,
as heuristic arguments suggest.
Fixed a function \ $\tomega(\tau)$  \ and  a 
`slow time' parameter $\varepsilon>0$, consider the family of equations
(\ref{eq1}) with \ $\omega(\ti;\varepsilon)=\tomega(\varepsilon \ti)$.
At least two precise senses are ascribed to the property of `adiabatic invariance' of $\I$:
\begin{enumerate} 

\item  For all $(q_0,p_0)\in\RR\times\RR^+$ and $T,\delta>0$ one can choose $\varepsilon>0$ so small that
 for all \  $ \ti \in[0,T/\varepsilon]$ \ the solution of (\ref{eq1}) with initial conditions $\big(q(0;\varepsilon),\dot q(0;\varepsilon)\big)=(q_0,p_0)$ satisfies
\be
|\I(\ti;\varepsilon)-\I(0;\varepsilon)|<\delta.                       \label{AdiabaticInv}
\ee
Namely, if we slow down the rate of variation of $\omega$ proportionally to $\varepsilon$
and simultaneously dilate the time interval proportionally to $1/\varepsilon$, so that 
the $\tau$-interval  $[0,T]$ and thus the variation of  $\omega$ in $[0,T/\varepsilon]$ don't change, the corresponding variation of  $\I$ vanishes with $\varepsilon$. As known (see e.g. \cite{Arnold}), a sufficient condition for this is $\tomega\in C^2(\RR^+)$. 

\item If $\tomega\in  C^{k+1}(\RR)$  ($k\!\in\!\NN$) and $\frac{d^h\tomega}{d\tau^h}\in L^1(\RR)$ for all $h=1,...,k\!+\!1$ (whence $\tomega(\tau)$ has well-defined limits and  vanishing derivatives
as $\tau\!\to\! \pm\infty$),  then  every solution of  (\ref{eq1})  satisfies (see e.g. \cite{LandauLifschitz})
\be
|\I(\infty;\varepsilon)-\I(-\infty;\varepsilon)|= O(\varepsilon^k);           \label{AdiabaticInv-k}
\ee
if  $\frac{d\tomega}{d\tau}\in {\cal S}(\RR)$, i.e. is a Schwarz  function (in particular, with compact support), 
then  (\ref{AdiabaticInv-k}) holds for all $k\in\NN$, namely the lhs goes to zero with $\varepsilon$ faster
than any power.
\end{enumerate} 

\noindent
In the appendix we prove both the sufficient condition in 1. and the implications in 2. quite  fast via our approach (an example of its usefulness).

In section \ref{conclu} we briefly mention some applications where these properties are important.

\subsection{Asymptotic expansion in the slow-time parameter $\varepsilon$}
\label{AsymptSect}

Similarly,  from the  sequence $\{\psi^{(h)}\}_{h\in\NN_0}$ one can obtain 
the asymptotic expansion of $\psi,\I$ in the slow time parameter $\varepsilon$ introduced in  section \ref{AdiabSect}.  One can iteratively 
show\footnote{Applying the definitions and the sine-of-a-sum rule we decompose the square bracket in (\ref{inter1})   into the sum of two terms, one proportional to $\sin(2\sigma^{(j)})=O\big(\sigma^{(j)}\big)$ and another to 
$\big[\cos(2\sigma^{(j)})\!-\!1\big]=O\Big(\big(\sigma^{(j)}\big)^2\Big)$; therefore it is a $O\big(\varepsilon^j\big)$ by
the induction hypothesis. As this is multiplied by 
$\dot\omega/\omega=\varepsilon d\tomega/d\tau$ the integrand is thus 
$O\big(\varepsilon^{j+1}\big)$, leading to (\ref{inter2}a). The proof of (\ref{inter2}b) is analogous.}
 that 
\be
\sigma^{(k)}=O(\varepsilon^k), \qquad
\chi^{(h)}:=\psi^{(h)}- \psi^{(h-1)}=O(\varepsilon^h), \qquad\quad h=1,...,k-1. \label{inter2}
\ee
Therefore the decomposition of  $\psi=\psi^{(k-1)}\!+\!\sigma^{(k)}$ into a sum of $\psi^{(0)}$
and subsequent corrections
\be
\psi=
\psi^{(0)}+\chi^{(1)}+...+\chi^{(k-1)}+\sigma^{(k)} \label{inter3}
\ee
is automatically a decomposition into terms $O(1),O(\varepsilon),...,O(\varepsilon^k)$.  
Replacing in (\ref{closedform-I})  one obtains the expansion for $\I$ via the Taylor formula for the exponential.
Integrating by parts one can iteratively extract from each integral $\chi^{(h)}= \int^\ti_{\ti_*}\!\!dz \left\{\frac{\dot\omega }{\omega}\!\left[\sin\!\left(2\psi^{(h-1)}\right) \!-\!\sin\!\left(2\psi^{(h-2)}\right)\right]\!\right\}\!(z)$ the leading contribution as a more explicit function times $\varepsilon^h$, putting the rest in the remainder.  In particular, for $k=2$ we obtain 
(choosing $\ti_*=0$ and abbreviating $\psi_0=\psi(0)$, $\tilde\zeta:=\frac 1{\tomega^2}\frac {d\tomega}{d\tau}$)
\bea
 \psi(\ti)=\varphi(\ti)-\frac \varepsilon 4\left\{\cos\big[2\varphi(\ti)\big]\,\tilde\zeta(\varepsilon \ti)-\cos(2\psi_0)\,\tilde\zeta(0)\right\}\!+ O(\varepsilon^2), \label{inter4} \\[8pt]
\I(\ti)=\I_0\left\{1-\frac \varepsilon 2\left[\sin\big[2\varphi(\ti)\big]\,\tilde\zeta(\varepsilon \ti)-\sin(2\psi_0)\,\tilde\zeta(0)\right] \right\}+ O(\varepsilon^2).   \label{inter5}
\eea

\subsection{Upper and lower bounds on the solutions}
\label{bounds}

We can easily bound $\psi(\ti),\I(\ti),H(\ti),...$ in an interval $[\ti_{h},\ti_{h+1}]$ 
where $\omega(\ti)$ is monotone.

If  either $\dot \omega\ge 0$ and $h$ is even, or $\dot \omega\le 0$ and 
$h$ is odd, then  \ $\dot \omega\sin(2\psi)\ge 0$ and by eq
 (\ref{eq2}) $0\le\dot\psi\!-\!\omega\le (-1)^h\dot\omega/2\omega$;  integrating  in  $[\ti_{h},\ti]$ with the initial condition $\varphi(\ti_h)=\psi(\ti_h)$ we find
\bea
\varphi(\ti)\le\psi(\ti)\le \varphi(\ti)+(-1)^h\log\sqrt{\frac{\omega(\ti)}{\omega(\ti_h)}}
=:\varphi^{(1)}(\ti)                             \label{psi-bounds1}
\eea  
for all $ \ti\in [\ti_{h},\ti_{h+1}]$.
In particular, choosing $\ti=\ti_{h+1}$ we find
\be
0\le\frac {\pi}2-\int^{\ti_{h+1}}_{\ti_h}\!\!\!\!\!\!dz\:\omega(z) \le (-1)^h\log\sqrt{\frac{\omega(\ti_{h+1})}{\omega(\ti_h)}}              \label{Deltay-bounds1}
\ee
which implicitly yield  bounds on the length \ $\ti_{h+1}\!-\!\ti_h$ \ of the
interval $[\ti_{h},\ti_{h+1}]$ that are more stringent than (\ref{Delta\ti_hbound}).
Moreover, let $\bar \ti_h\in [\ti_{h},\ti_{h+1}]$ be defined by the condition 
$\varphi^{(1)}(\bar \ti_{h+1})= (h\!+\!1)\pi/2$.
By (\ref{psi-specialvalues}), $\cos(2w)$  decreases, grows with $w\in[\psi(\ti_{h}),\psi(\ti_{h+1})]=\left[\frac{h\pi}2,\frac{(h\!+\!1)\pi}2\right]$  if $h$ is even, odd respectively.
This and  (\ref{psi-bounds1}) imply  in either case
\bea
-\frac{\dot\omega }{\omega}\cos(2\varphi)\le -\frac{\dot\omega }{\omega}\cos(2\psi)\le 
-\frac{\dot\omega }{\omega}\cos\left(2\varphi^{(1)}\right),         \label{interm0}
\eea
which replaced in (\ref{closedform-I}) give 
\bea
\exp\left\{-\!\!\int^\ti_{\ti_h}\!\!dz \left[\frac{\dot\omega }{\omega}\cos(2\varphi)\right]\!(z)\right\} 
\le \frac{\I(\ti)}{\I(\ti_h)}\le
\exp\left\{-\!\!\int^\ti_{\ti_h}\!\!dz \left[\frac{\dot\omega }{\omega}\cos\left(2\varphi^{(1)}\right)\right]\!(z)\right\};          \label{I-bounds1}
\eea
the left bounds in (\ref{interm0}-\ref{I-bounds1}) hold for all $ \ti\in [\ti_{h},\ti_{h+1}]$, while the right ones
only for $ \ti\in [\ti_{h},\bar \ti_{h+1}]$.

Similarly, if  $\dot \omega\!\le\! 0$ and $h$ is even, or $\dot \omega\!\ge\! 0$ and 
$h$ is odd, then  \ $\dot \omega\sin(2\psi)\!\le\! 0$  and by eq.
 (\ref{eq2}) $(-1)^h\dot\omega/2\omega\le\dot\psi\!-\!\omega\le 0$;  integrating  in  $[\ti_{h},\ti]$ with the initial condition $\varphi(\ti_h)=\psi(\ti_h)$ we find
\bea
\varphi^{(1)}(\ti)   :=\varphi(\ti)+(-1)^h\log\sqrt{\frac{\omega(\ti)}{\omega(\ti_h)}}\le\psi(\ti)\le \varphi(\ti)                                     \label{psi-bounds2}
\eea
 for all $ \ti\in [\ti_{h},\ti_{h+1}]$. In particular, choosing $\ti=\ti_{h+1}$ we find
\be
(-1)^h\log\sqrt{\frac{\omega(\ti_{h+1})}{\omega(\ti_h)}}\le\frac {\pi}2-\int^{\ti_{h+1}}_{\ti_h}\!\!\!\!\!\!dz\:\omega(z) \le 0                  \label{Deltay-bounds2}
\ee
which implicitly yield more stringent bounds on \ $\ti_{h+1}\!-\!\ti_h$ \ than  (\ref{Delta\ti_hbound}).
Moreover, let $\bar \ti_{h+1}'\in [\ti_{h},\ti_{h+1}]$ be defined by the condition 
$\varphi(\bar \ti_{h+1}')= (h\!+\!1)\pi/2$. Eq. (\ref{psi-bounds2}) implies  in either case again (\ref{interm0}) and (\ref{I-bounds1}); the only difference is that now 
the right bounds hold for all $ \ti\in [\ti_{h},\ti_{h+1}]$, while the left ones
only for $ \ti\in [\ti_{h},\bar \ti_{h+1}']$. 

{}From  (\ref{PolarCoord}), (\ref{I-bounds1}) and (\ref{psi-bounds1})  or (\ref{psi-bounds2}) one can now easily
derive bounds on $q,p$.

If $\omega(\ti)$ is monotone in a larger interval  $[\ti_{h},\ti_{h+k}]$ 
one can obtain there upper and lower bounds for $\psi(\ti)\!-\!\psi(\ti_h)$ and the ratio $\I(\ti)/\I(\ti_h)$ by using  (\ref{psi-bounds1}), (\ref{I-bounds1}) recursively in adjacent intervals. From the previous formulae and (\ref{Action-angle}),  (\ref{PolarCoord}) 
one  obtains corresponding bounds for $H(\ti),q(\ti),p(\ti)$.   The derivative  \ $\dot H=\dot\omega\omega q^2=2\dot\omega\I\sin^2\psi$ \  
of $H$ along  the solutions of  (\ref{eq1}) has the same sign of $\dot\omega$; therefore
$H(\ti)$ grows, decreases where $\omega(\ti)$ does.

One could obtain  upper and lower 
bounds also applying Liapunov's direct method to the family of Liapunov functions \
$V(\ti,\bar\omega):=\dot q{}^2(\ti)+\bar\omega^2 q^2(\ti)$ \ parametrized by
a positive constant $\bar\omega$. 
In each candidate interval $]\ti_h,\ti_{h+1}[$ one makes two choices of $\bar\omega$,  so as to make the derivative \
$\dot V=2\dot q q(\bar\omega^2\!-\!\omega^2)$ \ of $V$ along a solution $q(\ti)$
once positive and the other negative. Thus one determines upper and lower bounds first for $V$, then also for $q,\dot q$. However, it turns out that these bounds are rather less stringent and manageable than the ones 
found above.

\subsection{Parametric resonance, beats, damping}
\label{ParametricRes}

 Let us apply the previous results to Hill equation, i.e. (\ref{eq1}) with a periodic 
 $\omega$, \ $\omega(\ti+T)=\omega(\ti)$. \
We recall that, by Floquet theorem, the fundamental matrix solution   (\ref{matrix-sol})  of  $\dot V=AV$ fulfills
\be
V(\ti+nT)=V(\ti)M^n, \qquad M:=V(T)
\ee
 for all $\ti\in[0,T]$ and $n\in\NN$,  so  it suffices to determine $V$ in $\in[0,T]$.
As $\det M=1$, the eigenvalues of the monodromy matrix $M$ are $\lambda_\pm=\mu\pm\sqrt{\mu^2\!-\!1}$,
where $2\mu:=\mbox{Tr}(M)$, hence fulfill $\lambda_+\lambda_-=1$. Consequently
the two eigenvalues must do one of the following: i) be complex conjugate and of modulus 1 if $|\mu|<1$;
coincide if $|\mu|=1$; be both real one larger, and one smaller than 1, if  $|\mu|>1$. 
The trivial solution of (\ref{eq1})  is   stable if $|\mu|<1$, unstable if $|\mu|>1$ (most initial conditions near the trivial one   yield {\it parametrically resonant} solutions), see e.g. \cite{Arnold,LandauLifschitz}.
In terms of the solutions $\psi_a$ of (\ref{eq2}a) fulfilling $\psi_{1}(0)=\frac \pi 2$, $\psi_{2}(0)=0$, we find\footnote{
The $\psi_a$ associated to $\q_a$ ($a\!=\!1,2$) of (\ref{matrix-sol}) fulfill
$\psi_{1}(0)=\frac \pi 2$, $\I_{1}(0)=\frac {\omega_0}2$,   $\psi_{2}(0)=0$,  $\I_{2}(0)=\frac 1{2\omega_0}$. It follows
\bea
\q_1 =\sqrt{\frac{\omega_0}{\omega }} \sin(\psi_1)\, e^{-\Psi_1} ,\qquad
\dot\q_1 =\sqrt{\omega \omega_0} \cos(\psi_1)\, e^{-\Psi_1} , 
\nn
\q_2 =\frac1{\sqrt{\omega \omega_0}}  \sin(\psi_2)\, e^{-\Psi_2} ,\qquad
\dot\q_2 =\sqrt{\frac{\omega_0}{\omega }}\cos(\psi_2)\, e^{-\Psi_2} . \nonumber
\eea
where $\omega_0\!\equiv\!\omega(0)$. From $M=\left(\!\!\ba{ll}\q_1(T) &  \q_2(T)\\
\dot\q_1(T) & \dot\q_2(T)\ea\!\!\right)$ and $2\mu=\mbox{Tr}(M)=\q_1(T) +\dot\q_2(T)$ one  obtains (\ref{explicittrace}).
}
\bea
  2\mu
= e^{-\Psi_1(T)}\sin[\psi_1(T)]+e^{-\Psi_2(T)}\cos[\psi_2(T)], \label{explicittrace}
\eea
where $\Psi_a(\ti):=\!\!\int^\ti_0\!\!dz \left[\frac{\dot\omega }{2\omega}\cos(2\psi_a)\right]\!(z)$, \ $a=1,2$. If $\omega$ depends on a parameter $\eta$ so that \
$\dot\omega/\omega=O(\eta)$, \ then at  leading order in $\eta$ (i.e.,
up to corrections resp. 
of order 1,2,4) we find\footnote{In fact, according to (\ref{Inteq}b), 
$\psi_2^{(0)}(\ti)=\psi_1^{(0)}(\ti)\!-\!\frac\pi 2= \varphi(\ti)=\!\int^\ti_0\! dz \omega(z)=\bar\omega \ti+O(\eta)$.
}
\bea
\psi_2(\ti)=\psi_1(\ti)\!-\!\frac\pi 2= \bar\omega \ti ,\nn[8pt]
\Psi_2(\ti)= -\Psi_1(\ti)=\chi(\ti) \nn[8pt]
 \mu=\cos( \bar\omega T)\, \cosh\left[ \chi(T)\right],  \label{explicittrace-0}
\eea
where  the  characteristic (i.e. average) angular frequency $\bar\omega$  and $\chi(\ti)$ are defined by
\bea
\bar\omega:=\frac{\varphi(T)}T,\qquad
\chi(\ti)\!:=\!\!\int^\ti_0\! dz \left[\frac{\dot\omega }{2\omega}\cos(2\bar\omega \ti)\right]\!(z).
\nonumber
\eea
Clearly $\chi(T)=O(\eta)$.  We recover parametric resonance if the period $T$ 
of  variation of $\omega$ is one half of the characteristic
 time $2\pi/\bar\omega$, or  a multiple thereof, namely if   for some $j\in\NN$
\be
\bar\omega =j\, \frac{\pi}{T},                                        \label{resonancecond}
\ee
because then $\cos(\bar\omega T)=\pm 1$, and $|\mu|\simeq\cosh\left[ \chi(T)\right]>1$, 
unless $\chi(T)=0$. 
Evaluating  (\ref{explicittrace})  one can determine the region in parameter space $(\bar\omega,\eta)$ in the vicinity of a special value (\ref{resonancecond}) leading to stable or unstable solutions. At lowest order in 
 the small parameters \ $\eta$ and \ $b:=\bar\omega T\!-\!j\pi$ \ this amounts to evaluating  (\ref{explicittrace-0}). It is 
$\cos(\bar\omega T)=(-1)^j\cos b$, so that \
$2|\mu| -2=\left[ \chi(T)\right]^2-b^2 +O(\eta^4,b^4)$; \
therefore for sufficiently small $\eta,b$ the trivial solution will be stable if $|\chi(T)|<|b|$, 
parametrically resonant (hence unstable) if $|\chi(T)|>|b|$.

\begin{figure}[htbp]
 \includegraphics[width=7.3cm]{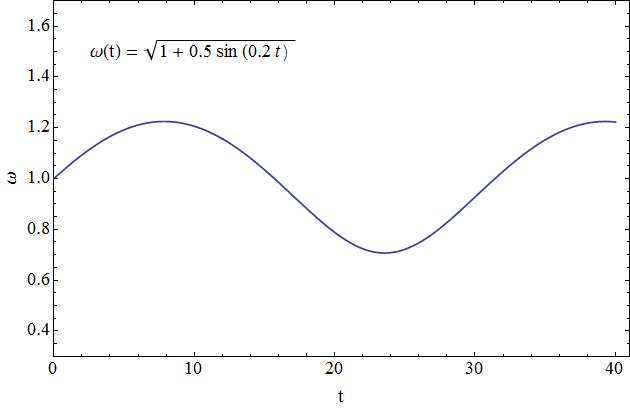}
\hskip1cm \includegraphics[width=7.3cm]{omega-Mathieu-nonresonant-epsilon02-a05}\\
 \includegraphics[width=8cm]{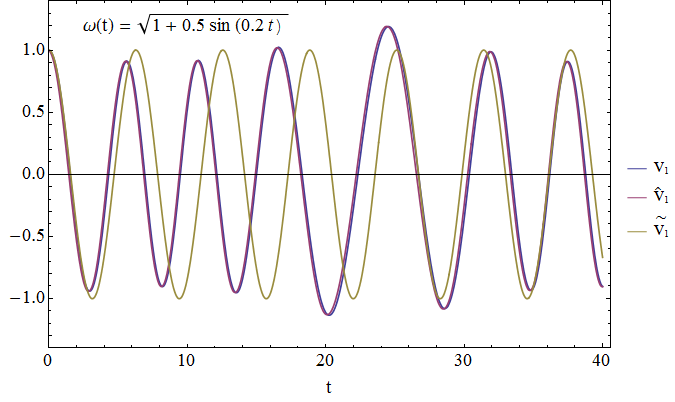}
\hfill \includegraphics[width=8cm]{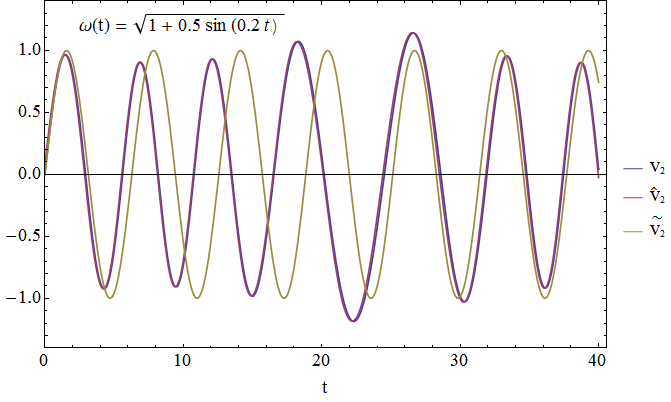}\\
  \includegraphics[width=8cm]{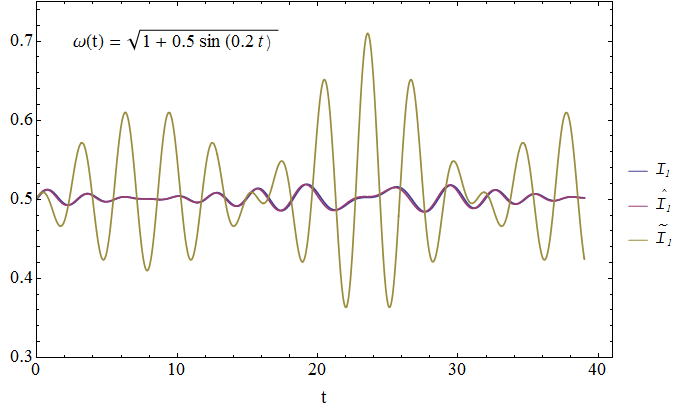}
\hfill \includegraphics[width=8cm]{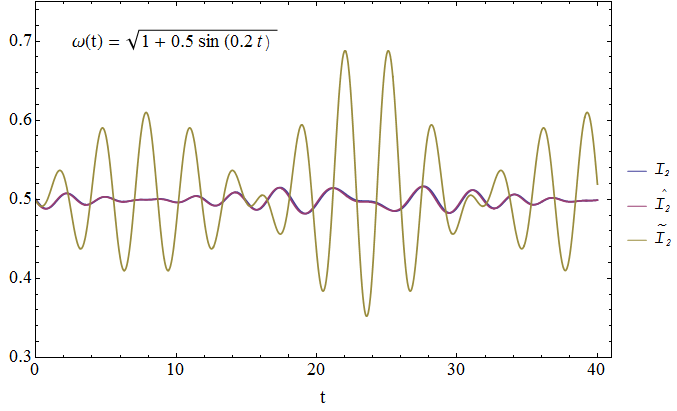}
 \caption{Mathieu equation with $\omega(\ti)=\bar\omega\sqrt{1+ 0.5\, \sin(0.2 \ti)}$ (this is a non-resonant case). Up: graph of $\omega$; center: graphs of the two solutions  $\q_a$ ($a=1,2$) of  (\ref{eq1}) fulfilling $\q_1(0)=\dot \q_2(0)=1$,
$\q_2(0)=\dot \q_1(0)=0$ (see section \ref{preli}) and of their approximations   $\hat \q_a,\tilde \q_a$, with $a=1$ on the left, $a=2$ on the right; down: graphs of the corresponding action variable  $\I_a$ and of  their approximations  $\hat \I_a,\tilde \I_a$. As we can see, $\hat \q_a\simeq \q_a$,  $\hat \I_a\simeq \I_a$, namely the approximation (\ref{def-hat-q}) is rather good,
and much better than the one   (\ref{tilde-approx}).}
\label{Mathieu-nonresonant-epsilon02-a05}
\end{figure}

\begin{figure}[htbp]
\hskip-.3cm  \includegraphics[width=7.6cm]{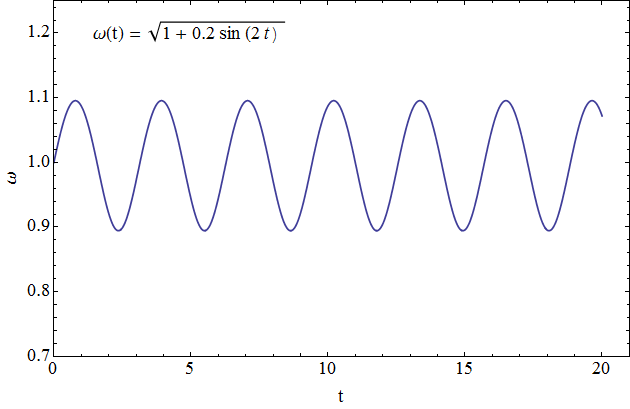}
\hskip1cm
 \includegraphics[width=7.6cm]{omega-Mathieu-resonant-epsilon2-a02}\\
\includegraphics[width=8cm]{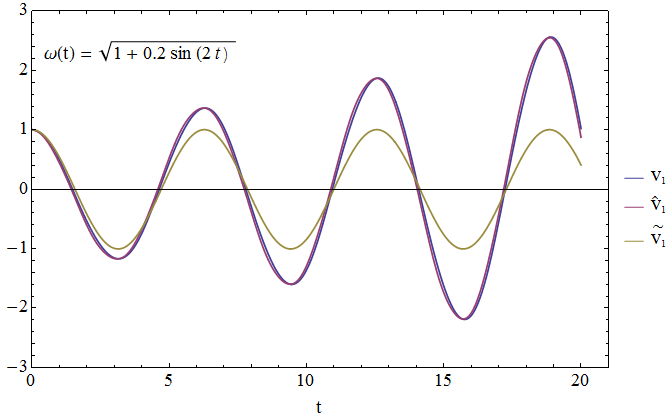}
\hfill
 \includegraphics[width=8cm]{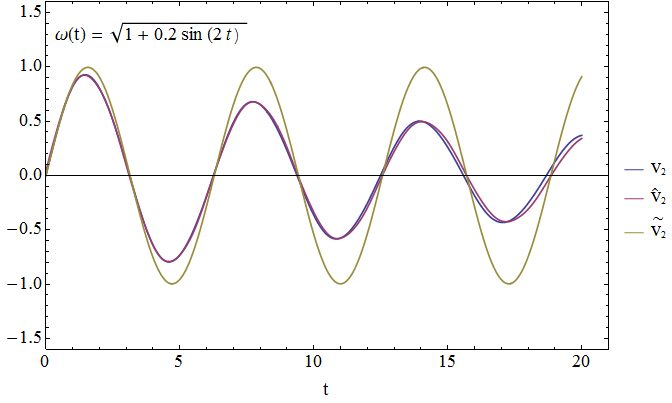}\\
.\hskip0.05cm   \includegraphics[width=7.9cm]{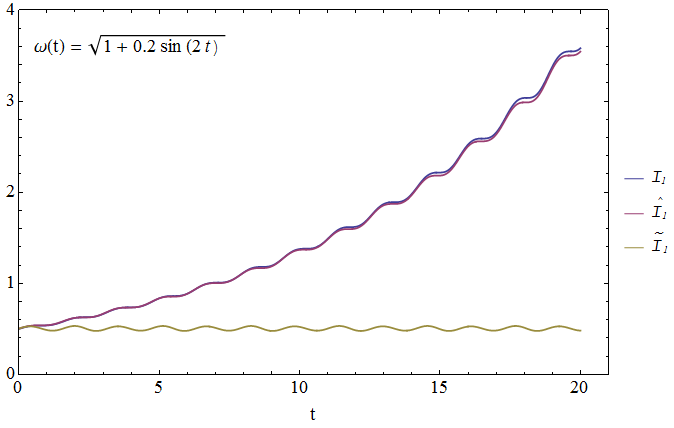}
\hskip0.5cm
 \includegraphics[width=7.9cm]{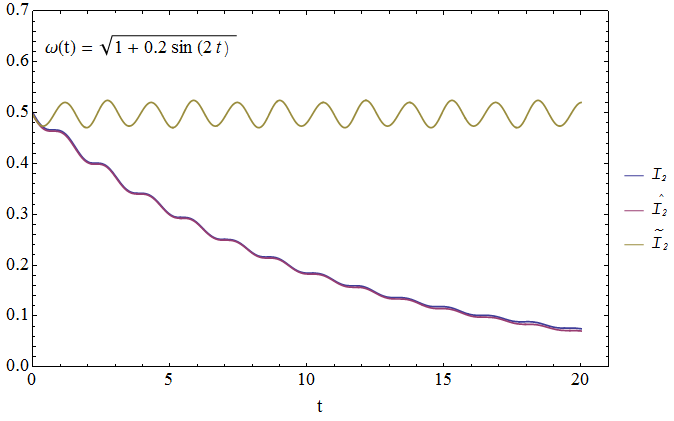}
 \caption{Mathieu equation with $\omega(\ti)=\bar\omega\sqrt{1+ 0.2\, \sin(2 \ti)}$
(this is a resonant case). Up: graph of $\omega$; center: graphs of the two solutions  $\q_a$ ($a=1,2$) of  (\ref{eq1}) fulfilling $\q_1(0)=\dot \q_2(0)=1$,
$\q_2(0)=\dot \q_1(0)=0$ (see section \ref{preli}) and of their approximations  $\hat \q_a,\tilde \q_a$, with $a=1$ on the left, $a=2$ on the right; down: graphs of the corresponding action variable  $\I_a$ and of  their approximations  $\hat \I_a,\tilde \I_a$.
Again, as we can see, $\hat \q_a\simeq \q_a$,  $\hat \I_a\simeq \I_a$, namely the approximation (\ref{def-hat-q}) is rather good,
and much better than the one   (\ref{tilde-approx}).}
\label{Mathieu-resonant-epsilon2-a02}
\end{figure}

\medskip
As an illustration, we determine at leading order
in $|\eta|<1$ the solutions of (\ref{eq2}) and $\mu$ if
\be
\omega(\ti)=\bar\omega\sqrt{1+ \eta \sin(\alpha \ti)},
\ee
with some constant $\alpha,\bar\omega>0$;  
the corresponding (\ref{eq1}) is the well-known Mathieu equation (which
rules many different phenomena, e.g. the forced motion of a swing, the
stability of ships,  the behaviour of parametric amplifiers based on electronic \cite{Howsmi70} or superconducting devices \cite{Lik70}, parametrically excited oscillations in microelectromechanical systems; see e.g. \cite{TurEtAl98} and references therein), and 
\bea
T=\frac{2\pi}{\alpha},\qquad 
\frac{\dot\omega }{\omega}=\frac{\eta \alpha\cos(\alpha \ti) }{2[1\!+\!\eta  \sin(\alpha \ti)]}
\nonumber
\eea
Denoting $\psi(0)=\psi_*$, $\I(0)=\I_*$, we find \ $\psi(\ti)=\psi^{(0)}(\ti)\!+\!O(\eta)=\psi_*\!+\!\bar\omega \ti\!+\!O(\eta)$, whence
\bea
\left[\frac{\dot\omega }{\omega}\cos\left(2\psi\right)\right]\!(\ti)
&=&
\frac{\eta\alpha}2 \cos(\alpha \ti)\cos(2\psi_*\!+\!2\bar\omega \ti)+ O(\eta^2)\nn[8pt]
&=& \frac{\eta\alpha}4\big\{\cos[2\psi_*\!+\!(2\bar\omega\!-\!\alpha) \ti]+
\cos[2\psi_*\!+\!(2\bar\omega\!+\!\alpha) \ti]\big\}+ O(\eta^2)\nonumber
\eea
Replacing in (\ref{closedform-I})   we find up to $O(\eta^2)$ \ $\log[\I_*/\I(\ti)]= 2\chi(\ti)$, \ with
\bea 
\chi(\ti)
= \left\{\!\!\ba{ll} \displaystyle \frac{\eta}4\bar\omega\left\{\cos(2\psi_*)\, \ti
+\frac{\sin(2\psi_*\!+\!4\bar\omega \ti)-\sin(2\psi_*)}{4\bar\omega}
\right\}  \qquad &\mbox{if }\alpha= 2\bar\omega,\\[12pt]
\displaystyle \frac{\eta\alpha}8\left\{\!\frac{\sin[2\psi_*\!+\!(2\bar\omega\!-\!\alpha) \ti]}{2\bar\omega\!-\!\alpha}+\frac{\sin[2\psi_*\!+\!(2\bar\omega\!+\!\alpha) \ti]}{2\bar\omega\!+\!\alpha}-\frac{4\bar\omega\sin(2\psi_*)}{4\bar\omega^2\!-\!\alpha^2}\!
\right\} \quad &\mbox{if }\alpha\neq 2\bar\omega.
\ea\right.
\nonumber
\eea
This leads to the following long-time behaviour.
If $\alpha =2\bar\omega$ we find \
$\I(\ti)\sim \exp[\eta\bar\omega\cos(2\psi_*) \ti/2]$, \ which exponentially
vanishes with $\ti$ if $\eta\cos(2\psi_*)<0$ (damping)
or diverges with $\ti$ if $\eta\cos(2\psi_*)>0$ (parametric resonance), respectively.
Otherwise $\log \I(\ti)$ is the superposition of two sinusoids. In particular, for $\alpha \simeq 2\bar\omega$ the behaviour of $\log \I(\ti)$ is a typical beat: $\log[\I(\ti)/\I(0)]$ oscillates approximately between  
$\pm|\eta|\alpha/4|2\bar\omega\!-\!\alpha|$ 
with a period $T=2\pi/(2\bar\omega\!-\!\alpha)$. In fig.s \ref{Mathieu-nonresonant-epsilon02-a05}, \ref{Mathieu-resonant-epsilon2-a02} we
have plotted (in a short interval starting from $\ti=0$) the exact fundamental solutions $\q_a$ introduced in section \ref{preli} and the corresponding action variables  $\I_a$ ($a=1,2$), as well as their lowest approximations [in the sense of (\ref{tilde-approx}), (\ref{def-hat-q})]
$\tilde\q_a,\tilde\I_a$, $\hat\q_a,\hat\I_a$  for a non-resonant and a resonant choice of the parameters $\eta,\alpha$. 
As evident, our approximations $\hat\q_a,\hat\I_a$ fit the exact solutions much better than the ones $\tilde\q_a,\tilde\I_a$. 

$\chi(T)$ is obtained setting \ $\psi_*=0$, $\ti=T$. \ At leading order in  $\eta$ we find
\bea
 \chi(T) 
= \left\{\!\!\ba{ll} \displaystyle  \eta\,\frac{\pi }4   \qquad &\mbox{if }\alpha= 2\bar\omega,\\[8pt]
\displaystyle 
\frac{\eta\alpha\bar\omega}{2[4\bar\omega^2\!-\!\alpha^2]}\sin\left(\!\frac{4\pi\bar\omega}{\alpha}\!\right)
 \quad &\mbox{if }\alpha\neq 2\bar\omega.
\ea\right.\nonumber
\eea
Near the resonance point $T =\pi/\bar\omega$, i.e. for small   \ $b=\bar\omega T\!-\!\pi\equiv\pi(2\bar\omega/\alpha\!-\!1)$, \
 we find \ $\chi(T) = \eta[\pi/4\!+\! O(b)]$ \ at lowest order in $b,\eta$.
 Therefore  near $(\bar\omega,\eta)=(\alpha /2,0)$ the regions fulfilling the inequalities \ $\pi|\eta|<4|b|$ \ and  \ $\pi|\eta|>4|b|$, \ or equivalently
$$
|\eta|< 4\left|\frac{2\bar\omega}{\alpha}\!-\!1\right|\qquad\mbox{and}\qquad |\eta|> 4\left|\frac{2\bar\omega}{\alpha}\!-\!1\right|,
$$
are respectively inside the stability and instability region, as known (see e.g. \cite{LandauLifschitz}).
Near the resonance points \ $T =j\,\frac{\pi}{\bar\omega}$ ($j>1$) \ one can determine the (in)stability regions by a more precise evaluation of  (\ref{explicittrace})\footnote{In fact,
for small  \ $b=\bar\omega T\!-\!j\pi\equiv\pi(2\bar\omega/\alpha\!-\!j)$ \ the previous formulae
give $\chi(T) =\eta b\frac{j}{j^2\!-\!1 }\!+\! O(b^2)$; \ 
 for small $|\eta|$ the inequality  $|\chi(T)|<|b|$ (whence $|\mu^{(0)}|<1$) is fulfilled for (small) $|b|\neq 0$,
while for $b=0$ formula (\ref{explicittrace-0}) gives $|\mu^{(0)}|=1$.}, which is out of the scope of this work.

\section{Discussion and conclusions}
\label{conclu}

In this paper we have reduced  (section \ref{action-angle}) - as far as we know for the first time - the integration of  (\ref{eq1}), 
or of more general linear systems (\ref{eq1'}a)  (section \ref{preli}),
to that of  the first order ODE  (\ref{eq2}a), or equivalently the integral equation  (\ref{Inteq}),
in  the  globally defined unknown phase $\psi$  
 and adopted the iterative resolution procedure (\ref{psi-h}) to obtain sequences $\{\psi^{(h)}\}_{h\in\NN_0}$
 {\it converging  uniformly} (and quite rapidly) to the solutions $\psi(\ti)$  in every compact interval
where $\omega$ is differentiable; $\psi(\ti)$ can be extended beyond discontinuity points of $\omega(\ti)$, if any, by
suitable matching conditions. 
Numerically, eqs. (\ref{psi-h}), (\ref{psi-h+}), (\ref{inter4}), and (\ref{inter5}) are the primary results to be  evaluated. Comparison with the numerical results in other approach will help in the validation of the method.
As a preliminary step we have studied (section \ref{zeroes}) the instants $\ti_h$ where $\psi$ is a multiple of $\pi/2$
(these are the interlacing zeros of $q(\ti),\dot q(\ti)$). 
The applications  sketched in the previous section \ref{Useful} illustrate hopefully  in a convincing way 
that our  approach is economical and effective, and that its interplay with the existing wisdom and alternative
approaches is rather promising for improving (both analitically and numerically) our knowledge 
of the solutions of (\ref{eq1}) and of various delicate aspects of theirs.
To that end, a comparison in particular  with the Ermakov reformulation  \cite{Ermakov}  
seems appropriate now. 

\noindent
Looking for $ q$ in the form of the product of an amplitude $\rho(\ti)$ and the sine of a phase 
$\theta(\ti)$,
\be
 q(\ti)=\rho(\ti)\sin[\theta(\ti)],                            \label{ansatz}
\ee
one easily finds that a sufficient condition for this to be a solution of  (\ref{eq1}) is that
$\rho,\theta$ fulfill the ODEs
\be
\rho\ddot\theta+2\dot\rho\dot\theta=0,\qquad \qquad
\ddot\rho=-\omega^2\rho+ \rho\dot\theta^2,           \label{ermakov}
\ee
which make the coefficients of $\sin\theta,\cos\theta$
vanish {\it separately}. By (\ref{ermakov}a),  $\rho^2\dot\theta$ has zero derivative and hence is a constant $L$.  Replacing $\rho^2\dot\theta\equiv L$ in (\ref{ermakov}b)  we arrive at 
the Ermakov equation \cite{Ermakov} in the unknown $\rho$
\be
\ddot\rho=-\omega^2\rho + \frac{L^2}{\rho^3}.                       \label{eqs1''}
\ee
Note that: i) the last term prevents $\rho$ to vanish anywhere; ii) the equation makes sense and yields solutions having continuous $\rho(\ti),\dot \rho(\ti)$ even if $\omega(\ti)$ is not. 

Given $L\neq 0$ and a {\it particular} solution $\rho(\ti)$ of (\ref{eqs1''}), the corresponding $\theta(\ti)$ is found integrating $\dot\theta=L/\rho^2$. Actually the fulfillment of (\ref{ermakov}), or equivalently of (\ref{eqs1''}) and  
$\theta(\ti)=\int^t\!\!d\tau L/\rho^2(\tau)$, is a sufficient condition for
\be
q_{A,\alpha}(\ti):=A\,\rho(\ti)\sin[\theta(\ti)+\alpha], \qquad A,\alpha\in\RR     \label{GenSol}
\ee
to be the {\it  general solution} of   (\ref{eq1}). In fact,  fixed any conditions 
$\big(q(\ti_*),\dot q(\ti_*)\big)=(q_*,p_*)$, if we set
$\alpha=\left\{\cot^{-1}\!\left[(p_*\rho^2/q_*\!-\!\dot\rho\rho)/L\right]\!-\!\theta\right\}\!(\ti_*)$,
$A=q_*/[\rho\sin(\theta\!+\!\alpha)](\ti_*)$ then $q_{A,\alpha}$ satisfies them.
Conversely, assume $q, \rho$ solve (\ref{eq1}), (\ref{eqs1''}) with some  constant $L\neq 0$. Then \cite{Ermakov} 
\be
2I:=\left( q \dot\rho- \dot q\rho\right)^2+L^2\!\left(\frac { q}{\rho}\right)^2 ,   \label{eqs2}
\ee
is a constant because $\dot I=0$. \  Of course, this {\it exact invariant} $I$ (usually dubbed after the names of Ermakov, Pinney,
Courant, Snyder, Lewis, Riesenfeld, in various combinations) 
must not be confused with the adiabatic one $\I$. Only when $\omega=$const eq. (\ref{ermakov}) 
admits the constant solution $\rho=\sqrt{L/\omega}$, which, replaced in (\ref{eqs2}), 
gives $I=L\I$, namely coinciding $I,\I$, up to normalization.
In general, the value of $I$ can be determined replacing in (\ref{eqs2}) the initial conditions fulfilled by $q,\rho$. 
In principle (\ref{eqs2}) allows to determine $\rho$ from $ q$, and conversely, but not in closed form.
A solution $\rho$ of  (\ref{eqs1''}) can be expressed explicitly \cite{Pinney} in terms of {\it two} 
independent solutions $u,v$ of (\ref{eq1}) as follows:
\be
\rho(\ti)=\sqrt{u^2(\ti)+\frac{L^2}{w^2}v^2(\ti)}\equiv\sqrt{q_1^2(\ti)+q_2^2(\ti)};       \label{q_1q_2->rho}
\ee
here $w=u\dot v-\dot u v=$const$\neq 0$ is the Wronskian of $u,v$, and in the last expression we have renamed $q_1:=u$, $q_2:=vL/w$.
A nice way \cite{EliGra76} to derive and interpret 
these results is to note that the vector ${\bm \rho}\equiv(q_1,q_2)$ satisfies again  (\ref{eq1})
regarded as a vector equation:
\be
\ddot{\bm \rho}=-\omega^2(\ti){\bm \rho};                     \label{eqs1}
\ee
this is the equation of motion of a particle in a plane  under the action
of a  time-dependent, but {\it central} elastic force   (${\bm \rho}$ is its  position vector with respect to the center $O$). Therefore
the angular momentum  $L:={\bm \rho}\times\dot{\bm \rho}=q_1\dot q_2\!-\!q_2\dot q_1$
(which coincides with the Wronskian  of $q_1,q_2$)  is conserved. 
Decomposing ${\bm \rho}=(\rho\cos\theta,\rho\sin\theta)$,
it is immediate to check that (\ref{ermakov}) amount to (\ref{eqs1}) written in the polar coordinates $\rho,\theta$, \ and   $L=\rho^2\dot\theta$. Replacing $q\mapsto q_1=\rho\cos\theta$ or 
$q\mapsto q_2=\rho\sin\theta$ in (\ref{eqs2}) one immediately finds that $2I=L^2$.  If $q_1,q_2$ are proportional then $L\!=\!0$, $\theta\!=$const,
and the particle oscillates along a straight line passing through $O$; otherwise $L\neq 0$, and the particle goes around $O$. 
Finally, the invariant (\ref{eqs2}) 
is related also to the symmetries of the equation (\ref{eq1}) (see \cite{QinDav06,ZhaZha16} and references therein).
Probably  it is worth  underlining that in general
$\rho\neq\sqrt{2\I/\omega}$, $\theta\neq\psi$ and, as already noted, $\I\neq I$,
albeit (\ref{PolarCoord}a),  (\ref{ansatz}) look similar, and  
both  $\sqrt{2\I/\omega},\rho(\ti)$ - contrary to $q(\ti)$ - keep their sign, so that 
they can play the role of modulating amplitudes (envelopes) for the solutions of (\ref{eq1})
resp. in (\ref{PolarCoord}), (\ref{ansatz}).

The exact invariant (\ref{eqs2})  is theoretically remarkable both in classical and quantum physics
[however, in the latter case $q,p$ are {\it operators}, while $\rho$ remains a {\it numerical} solution of  (\ref{eqs1''})].
For instance, in accelerator physics $I$ is a powerful constraint used
to characterize the motion of a charged particle in alternating-gradient field configurations \cite{CouSny58,DavQin01}.
In  quantum mechanics the eigenvectors of the {\it operator} $I$ have time-independent eigenvalues, 
make up an orthonormal basis of the Hilbert space of states and may be normalized so as to be solutions 
of the Schroedinger equation; they can be built using ladder operators $a,a^\dagger$
as in the $\omega=$const case, and $I=a^\dagger a$  \cite{Lew67,Lew68JMP,LewRies69} (see e.g. also \cite{Man96,FioGou11}). 
However, the concrete use of $I$
is based on the knowledge of a solution of  (\ref{eqs1''}) for the specific problem at hand,
but solving the second order ODE (\ref{eqs1''}) is not easier than solving the one  (\ref{eq1}), except for special cases, and in general is more difficult than solving the  first order ODE (\ref{eq2}a). Therefore our iterative resolution method (\ref{psi-h}) 
 could be used also for constructing  solutions $\rho$ of  (\ref{eqs1''})  via  (\ref{q_1q_2->rho}).

Finding the most general 
conditions for the asymptotic stability of the trivial solution when external forces  vanish and
only the damping depends on time is also of interest \cite{Hat18}.

Using our approach to determine the parametric resonance regions of  (\ref{eq1}) for specific classes of
$\omega(\ti)$ (see e.g. \cite{Nes13}) is certainly also worth investigation. 
This might be useful also for quantum harmonic oscillators. For instance, it would be interesting to relate our procedure with a recursive resolution procedure  \cite{MarEtAl20,MarEtAl20',MarPirFar21,MarEtAl23} that has bee recently applied to analyzing parametric resonance, as well as the effects of adiabatic or sudden frequency variations. 

We believe that these and many other applications of our method are at hand.

\appendix
\section{Appendix}
\label{App}
  \renewcommand{\theequation}{A.\arabic{equation}}
  \setcounter{equation}{0}  

\subsection{Proof of Proposition \ref{SpecialPoints} }

For all $\ti_*\!\in\!\RR$  it is $ (q_*,p_*)\neq (0,0)$, because $q(\ti)$ is a nontrivial solution.
If $q_*= 0$ then respectively set $\ti_0:=\ti_*$  if $p_*>0$,   $\ti_2:=\ti_*$ if $p_*<0$.
If $q_*\neq 0$ then $r_*\in\RR$; in the  largest  interval containing $\ti_*$ where $q(\ti)$
keeps its sign we find $-\dot r\ge \omega_l^2+r^2$, whence
\bea
\frac{-\dot r/\omega_l}{1+(r/\omega_l)^2}
=\frac d{d\ti} \cot^{-1}\!\left(\!\frac r{\omega_l}\!\right)\ge \omega_l \quad\Rightarrow\quad	\cot^{-1}\!\left[\frac {r(\ti)}{\omega_l}\right]\ge \beta_*\!+\! \omega_l (\ti\!-\!\ti_*)=:\varphi_l(\ti).
\label{r-arg}
\eea
where $\beta_*\equiv\cot^{-1}\!\left(\frac {r_*}{\omega_l}\right)\!\in\,]0,\pi[$.
Let $\bar \ti_\pm
$ be defined by the conditions 
$\varphi_l(\bar \ti_-)=0$, 
$\varphi_l(\bar \ti_+)=\pi$; 
by construction $
\ti_*\in]\bar \ti_-,\bar \ti_+[$. 
The last inequality implies
$r(\ti)/\omega_l   \le \cot\left[\varphi_l(\ti)\right]$; since the rhs goes to $-\infty$ as $\ti\uparrow \bar \ti_+$, there must exist a $ \ti_+\!\in\,] \ti_*,\bar \ti_+[$  such that $r(\ti)\to-\infty$ as
$\ti\uparrow  \ti_+$. If  $p(\ti)>0$, $q(\ti)<0$ (resp. $p(\ti)<0$, $q(\ti)>0$) in a left neighbourhood of $\ti_+$ then the point $\ti_0:=\ti_+$ (resp. $\ti_2:=\ti_+$) has indeed the property mentioned in the claim.

In either case, now set $\ti_*\equiv \ti_0$ (resp. $\ti_*\equiv \ti_2$), $u\equiv\omega_ls$. In the largest interval $]\ti_*,\ti_+'[$  where $p(\ti)$ keeps its sign
we find $\dot s\ge 1+u^2$, whence, integrating over $[\ti_*,\ti]$,
\bea
\frac{\dot u}{1\!+\!u^2}
=\frac d{d\ti} \tan^{-1}\! u\ge \omega_l \quad\Rightarrow\quad	\tan^{-1}\!\left[u(\ti)\right]\ge 
\omega_l (\ti\!-\!\ti_*)
\quad\Rightarrow\quad	\omega_l  s(\ti)\ge
\tan[\omega_l (\ti\!-\!\ti_*)] \label{s-arg}
\eea
Since the rhs diverges as $\ti\uparrow \ti_*\!+\!\pi/2\omega_l$, it must be $ \ti_+'\!\in\,]\ti_*,\ti_*\!+\!\pi/2\omega_l[$ and $s(\ti)\to+\infty$ as
$\ti\uparrow  \ti_+'$; the point $\ti_1:=\ti_+'$ (resp. $\ti_3:=\ti_+'$) has indeed the property mentioned in the claim.

Setting $\ti_*=\ti_1$  (resp. $\ti_*=\ti_3$) and  using again (\ref{r-arg}), with $\beta_*=\pi/2$,
we prove the existence of $\ti_2$  (resp. $\ti_4$); setting $\ti_*=\ti_2$  (resp. $\ti_*=\ti_4$) and  using again (\ref{s-arg})
we prove the existence of $\ti_3$  (resp. $\ti_5$); and so on.
Replacing $\ti\mapsto -\ti$ in the equation and using the previous results
one iteratively proves the existence of the $\ti_h$ for $h\!\in\!\ZZ$ going to $-\infty$. 

Finally, if $\omega(\ti)\ge\bar\omega_l>0$ for all $\ti\in J\subset \RR$ 
the claim follows from the previous case after extending $\omega(\ti)$ so that
$\omega(\ti)\ge\bar\omega_l$ for all $\ti\in \RR$.

\subsection{Proof of  Proposition \ref{MonotonSpecialPoints}}

\begin{lemma}
Let   $ q_1, q_2$ be two independent solutions  of (\ref{eq1}) having  initial data at $\ti_*$ in the same quadrant of the $(q,p)$ plane, and let $\hat \ti_1>\ti_*$, $\hat \ti_2>\ti_*$
be the points where  $ q_1, q_2$ change quadrant,  respectively.
Then $\hat \ti_1< \hat \ti_2$ provided the initial data of both $ q_1, q_2$ at $\ti_*$ fulfill either \
$\dot  q_{1*} q_{2*}<\dot  q_{2*} q_{1*}\le 0$, \ or 
 \ $0\le\dot  q_{1*} q_{2*}<\dot  q_{2*} q_{1*}$.
\label{CompareZeroes}
\end{lemma}

\bp{}  The difference of  (\ref{rseqs}a) for the corresponding ratios  $r_1=\dot q_1/ q_1,r_2=\dot q_2/ q_2 $ yields $\dot r_1\!-\!\dot r_2=-(r_1\!+\!r_2)(r_1\!-\!r_2)$.
If the initial data of  $ q_1, q_2$ at $\ti_*$ fulfill $\dot  q_{1*} q_{2*}<\dot  q_{2*} q_{1*}\le 0$ (so that they both belong either to the first or to the third quadrant), then  $r_1(\ti_*)<r_2(\ti_*)\le 0$, $\dot r_1\!-\!\dot r_2$ is  negative at $\ti=\ti_*$, keeps negative for all 
$\ti\in[\ti_*,\ti''[$, and
so does $r_1\!-\!r_2$; here 
$\ti''\!\equiv\!\min\{\hat \ti_1, \hat \ti_2\}$.
Hence $r_2$ cannot diverge before $r_1$, i.e.  $\ti''=\hat \ti_1\le \hat \ti_2$. 
Dividing by $(r_1\!-\!r_2)$
we find that 
$$
\frac d {dy}\log(r_2-r_1)=-(r_1\!+\!r_2)\stackrel{\ti\to \hat \ti_1}{-\!\!\!-\!\!\!\longrightarrow}+\infty
$$
and this excludes that $r_2$ diverges together with $r_1$, i.e. we find $\hat \ti_1< \hat \ti_2$,
as claimed. 

Similarly, the difference of  (\ref{rseqs}b) for the  ratios
  $s_1= q_1/\dot q_1,s_2=  q_2/\dot q_2$  
yields $\dot s_1\!-\!\dot s_2=\omega^2(s_1\!+\!s_2)(s_1\!-\!s_2)$.
If the initial data of $ q_1, q_2$ at $\ti_*$ fulfill 
$0\le\dot  q_{1*} q_{2*}<\dot  q_{2*} q_{1*}$  (so that they both belong either to the second or to the fourth quadrant)
then  $s_1(\ti_*)>s_2(\ti_*)\ge 0$, whence, arguing as before, $\hat \ti_1<\hat \ti_2$. 
\ep


\bigskip
\noindent 
To show   that $\ti_i'>\ti_i$ implies  $\ti_h(\ti_i')>\ti_h(\ti_i)$ for all $h\in\ZZ$
we apply the lemma setting \ $ q_1(\ti)\!\equiv\! Q(\ti;\ti_i,q_i,p_i)$,  $ q_2(\ti)\!\equiv\!Q(\ti;\ti_i',q_i,p_i)$.

\medskip
We first consider the case that $q_i\le 0$, $\dot q_i>0$ (fourth quadrant). Let us denote by $\ti_0(\ti_i)$
the smallest $\ti\ge\ti_i$ such that  $q_1(\ti)=0$ and by  $\ti_0(\ti_i')$ the smallest $\ti\ge\ti_i'$ such that  $q_2(\ti)=0$.
In particular, if $q_i= 0$,  then  $\ti_0(\ti_i)=\ti_i>\ti_i'=\ti_0(\ti_i')$. If $q_i< 0$
and $ \ti_i' \ge \ti_0(\ti_i)$,  then a fortiori $\ti_0(\ti_i')>  \ti_i' \ge \ti_0(\ti_i)$; \ if $q_i< 0$
and $\ti_i'\in]\ti_i,\ti_0(\ti_i)[$, setting $\ti_*\equiv  \ti_0(\ti_i)$,
we find  $ 0=\dot  q_{2*} q_{1*} > \dot  q_{1*} q_{2*} $
and, applying the previous lemma, $\ti_0(\ti_i)=\hat \ti_1<\hat \ti_2=\ti_0(\ti_i')$. Namely, 
in all cases we find  $\ti_0(\ti_i')>  \ti_0(\ti_i)$, as claimed.

If $\ti_0(\ti_i')\ge \ti_1(\ti_i)$ then a fortiori $\ti_1(\ti_i')>  \ti_1(\ti_i)$; \ if $\ti_0(\ti_i')\in]\ti_0(\ti_i), \ti_1(\ti_i)[$,
then setting $\ti_*\equiv \ti_0(\ti_i')$ we find   $0=\dot  q_{1*} q_{2*}<\dot  q_{2*} q_{1*}$ 
and, by the previous lemma, 
$\ti_1(\ti_i)=\hat \ti_1<\hat \ti_2=\ti_1(\ti_
i')$; namely, 
in both cases we find  $\ti_1(\ti_i')>  \ti_1(\ti_i)$, as claimed. 

If $\ti_1(\ti_i')\ge \ti_2(\ti_i)$ then a fortiori $\ti_2(\ti_i')>  \ti_2(\ti_i)$; \ if $\ti_1(\ti_i')\in]\ti_1(\ti_i), \ti_2(\ti_i)[$,
then setting $\ti_*\equiv \ti_1(\ti_i')$ we find   $\dot  q_{1*} q_{2*}<\dot  q_{2*} q_{1*}= 0$ 
and, by the previous lemma, 
$\ti_2(\ti_i)=\hat \ti_1<\hat \ti_2=\ti_2(\ti_
i')$; namely, 
in both cases we find  $\ti_2(\ti_i')>  \ti_2(\ti_i)$, as claimed. 

If $\ti_2(\ti_i')\ge \ti_3(\ti_i)$, then a fortiori $\ti_3(\ti_i')>  \ti_3(\ti_i)$; \
if $\ti_2(\ti_i')\in]\ti_2(\ti_i), \ti_3(\ti_i)[$,
then setting $\ti_*\equiv \ti_2(\ti_i')$,  we find again 
$ \dot  q_{2*} q_{1*} > \dot  q_{1*} q_{2*} =0$ and, by the previous lemma, 
$\ti_3(\ti_i)=\hat \ti_1<\hat \ti_2=\ti_3(\ti_i')$; namely, 
in both cases we find  $\ti_3(\ti_i')>  \ti_3(\ti_i)$, as claimed. And so on, for all nonnegative $h$.
The claim for negative $h$ follows after replacing $\ti\mapsto -\ti$.

\medskip
In the case that $q_i> 0$, $\dot q_i\ge0$ (first quadrant) we denote by $\ti_1(\ti_i)$
the smallest $\ti\ge\ti_i$ such that  $\dot q_1(\ti)=0$ and by  $\ti_1(\ti_i')$ the smallest $\ti\ge\ti_i'$ such that  $\dot q_2(\ti)=0$.
In particular, if $\dot q_i= 0$,  then  $\ti_1(\ti_i)=\ti_i>\ti_i'=\ti_1(\ti_i')$. If $\dot q_i>0$
and $ \ti_i' \ge \ti_1(\ti_i)$,  then a fortiori $\ti_1(\ti_i')>  \ti_i' \ge \ti_1(\ti_i)$; \ if  $\dot q_i>0$
and $\ti_i'\in]\ti_i,\ti_1(\ti_i)[$, setting $\ti_*\equiv  \ti_1(\ti_i)$,
we find   $0=\dot  q_{1*} q_{2*}<\dot  q_{2*} q_{1*}$ 
and, applying the previous lemma, $\ti_1(\ti_i)=\hat \ti_1<\hat \ti_2=\ti_1(\ti_i')$. Namely, 
in all cases we find  $\ti_1(\ti_i')>  \ti_1(\ti_i)$, as claimed.
The rest of the proof goes as above.

\medskip
In the case that $q_i\ge 0$, $\dot q_i <0$ (second quadrant) we denote by $\ti_2(\ti_i)$
the smallest $\ti\ge\ti_i$ such that  $q_1(\ti)=0$ and by  $\ti_2(\ti_i')$ the smallest $\ti\ge\ti_i'$ such that  $q_2(\ti)=0$.
In particular, if $q_i= 0$,  then  $\ti_2(\ti_i)=\ti_i>\ti_i'=\ti_2(\ti_i')$. If $q_i<0$
and $ \ti_i' \ge \ti_2(\ti_i)$,  then a fortiori $\ti_2(\ti_i')>  \ti_i' \ge \ti_2(\ti_i)$; \ if $q_i<0$
and $\ti_i'\in]\ti_i,\ti_2(\ti_i)[$, setting $\ti_*\equiv  \ti_2(\ti_i)$,
we find   $\dot  q_{1*} q_{2*}<\dot  q_{2*} q_{1*}= 0$ 
and, applying the previous lemma, $\ti_2(\ti_i)=\hat \ti_1<\hat \ti_2=\ti_2(\ti_i')$. Namely, 
in all cases we find  $\ti_2(\ti_i')>  \ti_2(\ti_i)$, as claimed.
The rest of the proof goes as above.

\medskip
In the case that $q_i<0$, $\dot q_i \le 0$ (third quadrant) we denote by $\ti_3(\ti_i)$
the smallest $\ti\ge\ti_i$ such that  $\dot q_1(\ti)=0$ and by  $\ti_3(\ti_i')$ the smallest $\ti\ge\ti_i'$ such that  $\dot q_2(\ti)=0$.
In particular, if $\dot q_i= 0$,  then  $\ti_3(\ti_i)=\ti_i>\ti_i'=\ti_3(\ti_i')$. If $\dot q_i<0$
and $ \ti_i' \ge \ti_3(\ti_i)$,  then a fortiori $\ti_3(\ti_i')>  \ti_i' \ge \ti_3(\ti_i)$; \ if $\dot q_i<0$
and $\ti_i'\in]\ti_i,\ti_3(\ti_i)[$, setting $\ti_*\equiv  \ti_3(\ti_i)$,
we find   $ \dot  q_{2*} q_{1*} > \dot  q_{1*} q_{2*} =0$
and, applying the previous lemma, $\ti_3(\ti_i)=\hat \ti_1<\hat \ti_2=\ti_3(\ti_i')$. Namely, 
in all cases we find  $\ti_3(\ti_i')>  \ti_3(\ti_i)$, as claimed.
The rest of the proof goes as above.

\subsection{Proof of the bounds (\ref{sigma_h-bound-0'})}

First  note that $f(\ti,u)$ is continuous in $\ti$ and uniformly Lipschitz continuous in $u$, since
\be
|f(\ti,u_1)-f(\ti,u_2)| \leq \frac{\mu}2 |u_1\!-\!u_2|,   \qquad \mu:=\sup_K   \Big|\frac{\dot\omega}{\omega}\Big|<\infty.          \label{Lipsch}
\ee
For all $\lambda\in\RR^+$ 
$C(K{})$ is a Banach space  w.r.t. the norm $\|\:\|_{\lambda, {\scriptscriptstyle K}}$, which
is equivalent to the one $\Vert\:\Vert_\infty$. \
Eq. (\ref{Lipsch}) implies \ $\left|f[\ti,u_1(\ti)]-f[\ti,u_2(\ti)]\right| e^{-\lambda |\ti-\ti_*|}\le (\mu/2) \Vert u_1\!-\!u_2\Vert_{\lambda, {\scriptscriptstyle K}}$, whence 
$$
\qquad\qquad  \|{\cal T}u_1\!-\!{\cal T}u_2\|_{\lambda, {\scriptscriptstyle K}} \: <\: \frac{\mu}{2\lambda} \: \|u_1\!-\!u_2\|_{\lambda, {\scriptscriptstyle K}}          
$$
for all $u_1,u_2\in C(K{})$. This follows  from the inequalities
\bea
\big| [{\cal T}\!u_1\!-\!{\cal T}\!u_2](\ti) \big| e^{\lambda (\ti_*-\ti)}\leq 
\int^\ti_{\ti_*}\!\!dz\: \big|f[z,\!u_1\!(z)]\!-\!f[z,\!u_2\!(z)]\big|e^{\lambda (\ti_*-\ti)} \nn
\leq \frac{\mu}{2} \|u_1\!-\!u_2\|_{\lambda, {\scriptscriptstyle K}} \!\! \int^\ti_{\ti_*}\!\!dz\:e^{\lambda (\ti_*-z)}
<\frac{\mu}{2\lambda} \: \|u_1\!-\!u_2\|_{\lambda, {\scriptscriptstyle K}}  \nonumber
\eea
if $\ti> \ti_*$, and from the ones with $\ti,\ti_*$ exchanged if $\ti< \ti_*$.
Hence $\T$ is a contraction of $C(K{})$ provided we choose $2\lambda\!>\!\mu$, and
Banach  fixed point theorem can be applied. Eq. (\ref{sigma_h-bound-0'}) follows from the byproduct of that theorem 
\be
 \left\|\psi^{(h)}\!-\!\psi\right\|_{\lambda, {\scriptscriptstyle K}} \: <\: \frac{\nu^h}{1\!-\!\nu} \: 
 \left\| \varphi\!-\!\psi^{(1)}\right\|_{\lambda, {\scriptscriptstyle K}},
 \qquad\quad  \mbox{where }\:\nu:= \frac{\mu}{2\lambda}\in]0,1[.
\label{sigma_h-bound-0}
\ee

\subsection{Proof of the adiabatic invariance properties (\ref{AdiabaticInv}),  (\ref{AdiabaticInv-k})}

\noindent
If we adopt $\tau =\varepsilon \ti$ as the `time' variable eq. (\ref{eq2}a) takes the form
\be
\frac{d\psi}{d\tau}= \frac{\tomega}{\varepsilon} +\frac1{2\tomega}\frac{d\tomega}{d\tau} \sin(2\psi).               \label{eq2a'}
\ee
The function \ $\varphi(\tau):=\varphi_0\!+\!\int^\tau_{0}\!\!d\tau'\,\tomega(\tau')$ is strictly growing
(here $\varphi_0\!\equiv\!\psi(0)$);
to shorten the proof we make the further change of independent variable $\tau\mapsto \varphi$. We denote a transformed function putting a hat above its symbol,  
abbreviate $\hat u'(\varphi)\equiv d\hat u/d\varphi$ and more generally $\hat u^{[h]}\equiv d^h\hat u/d\varphi^h$.
As $\varphi$ is dimensionless, not only
$\hv=\homega'/\homega
$, but also all  derivatives $\hv',\hv'',...$ are; moreover, $\homega(\varphi)$ fulfills the same conditions as $\tomega(\tau)$\footnote{This follows from expressing  $\frac d{d\varphi}=\frac 1{\tomega}\frac d{d\tau}$ and using the   bounds for 
$d\varphi/d\tau=\tomega(\tau)$ and its derivatives.}, namely
\bea
\ba{ll}
\sup\limits_{\tau\in[0,T]}\left|\frac{d^j\tomega}{d\tau^j}\right|<\infty,\quad c_h:=\sup\limits_{\alpha\in[\psi_0,\varphi(T)]}\left|\hv^{[h]}(\alpha)\right|<\infty, \quad  j=0,1,2,
\quad h=0,1,
\quad & \mbox{in case 1}, \\[12pt] 
\sup\limits_{\tau\in\RR}\!\left|\frac{d^j\tomega}{d\tau^j}\right|<\infty,\quad
\sup\limits_{\RR}\!\left|\hv^{[h]} \right|<\infty, \quad\hv^{[h]}\!\in\! L^1\!(\RR), 
\quad  j\!=\!0,\!...,k\!+\!1, \:\:  h\!=\!0,\!...,k
\quad & \mbox{in case 2}.
\ea \label{hv-bounds}
\eea   
Eq.  (\ref{eq2a'}),   (\ref{closedform-I}) take the form
\bea
\hpsi'=\frac{f}{\varepsilon},\qquad f :=1+\varepsilon\frac \hv2 \sin(2\hpsi) \label{eq2'}\\
-\log\frac{\hI(\varphi)}{\hI(\varphi_*)}=
\!\!\int^\varphi_{\varphi_*}\!\!d\beta \left[ \hv\,\cos\left(2\hpsi\right)\right]\!(\beta); \label{closedform-I'}
\eea
If $\tomega\in  C^{k+1}(\RR)$  then $\hv\in  C^k(\RR)$, and the integral at the rhs of (\ref{closedform-I'}) can be transformed integrating by parts $k$ times. The result can be extracted as the real part of
\bea
\int^\varphi_{\varphi_*}\!\!d\beta \left[ \hv\, e^{i2\hpsi}\right]\!(\beta)
&=& \int^\varphi_{\varphi_*}\!\!d\beta \: \frac{\varepsilon \hv(\beta)}{i2f(\beta) } \frac{ d}{d\beta}e^{i2\hpsi(\beta)}  \nn
&=& \left[e^{i2\hpsi}\:\frac{\varepsilon \hv}{i2f }\right]^{\beta=\varphi}_{\beta=\varphi_*}\!\!+
\int^\varphi_{\varphi_*}\!\!d\beta \left[ e^{i2\hpsi}\:  \frac d{d\beta}\!  \left(\frac{i\varepsilon \hv}{2f}\right)\!\right]\!(\beta)=... \nn
&=&  \left\{e^{i2\hpsi}\sum_{h=0}^{k-1}\left(\!\frac{i\varepsilon}{2f} \frac d{d\beta}\!\right)^h\!\frac {\varepsilon \hv}{i2f}\!\right\}^{\beta=\varphi}_{\beta=\varphi_*}\!\!+
\varepsilon^k\int\limits^\varphi_{\varphi_*}\!\!d\beta \left[ e^{i2\hpsi}
\left(\! \frac d{d\beta}\frac{i}{2f}\!\right)^{k}\!\!v\right]\!\!(\beta). \label{k-parts}
\eea
In particular, if $\tomega\in  C^{2}(\RR)$ then 
\bea
-\log\frac{\hI(\varphi)}{\hI(\varphi_0)} &=& \varepsilon\left[\sin(2\hpsi)\:\frac { \hv}{2f}\right]^\varphi_{\varphi_0}\!\!-\frac \varepsilon 2
 \int^\varphi_{\varphi_0}\!\!\!d\beta \left[ \sin(2\hpsi)\:\frac d{d\beta}\!  \left(\!\frac{\hv}{f}\!\right)\!\right]\!(\beta)\nn
&=& \varepsilon\left[\sin(2\hpsi)\:\frac { \hv}{2f}\right]^\varphi_{\varphi_0}\!\!-\frac \varepsilon 2
 \int^\varphi_{\varphi_0}\!\!\!d\beta \left[ \sin(2\hpsi)   \left(\frac{\hv'}{f^2}\!-\! \frac{\hv^2}{f}\cos(2\hpsi)\!\right)\!\right]\!(\beta),  \nonumber
\eea
whence, noting that \  $f>1\!-\!\varepsilon c_0/2$,
\bea
\left|\log\frac{\hI(\varphi)}{\hI(\varphi_0)}\right|
\le  \varepsilon\frac{\hv(\varphi)\!+\!\hv(0) }{2} + 
\frac \varepsilon 2
 \int^\varphi_{0}\!\!\!d\beta \left(  \left|\frac{\hv'}{f^2}\right|\!+\! \left|\frac{\hv^2}{f}\right|\right)\!(\beta) 
\le \varepsilon\left\{c_0\!+\!\frac{\varphi\!-\!0}{2(1\!-\!\varepsilon \frac{c_0}2)} \left[ \frac{c_1}{1\!-\!\varepsilon \frac{c_0}2}\!+\!c_0^2\right]\right\}, \nonumber
\eea
or, using again the time as the independent variable,
\bea
\left|\log\frac{\I(\ti)}{\I(0)}\right|
\le  \varepsilon\left\{c_0\!+\!\frac{T\omu}{2\!-\!\varepsilon c_0} \left[ \frac{c_1}{1\!-\!\varepsilon \frac{c_0}2}\!+\!c_0^2\right]\!\right\}=:M(\varepsilon), \quad\Rightarrow\quad
|\I(\ti)-\I(0)|<\I(0) \left[e^{M(\varepsilon)}\!-\!1\right].
\nonumber
\eea
Clearly $M(\varepsilon)\to 0$ as $\varepsilon\! \to\! 0$. Since $0\!<\! x\!\le\! 1$ implies 
$e^x\!-\!1\!<\!2x$,
choosing  $\varepsilon$ so small that $M(\varepsilon)\!<\!\min\big\{1,\delta/2\I(0)\big\}$ it follows $e^{M(\varepsilon)}\!-\!1\!<\!2 M(\varepsilon)
\!<\!\delta/\I(0)$, and   (\ref{AdiabaticInv}) is fulfilled, as claimed.
In case 2, by (\ref{hv-bounds}b) the terms  in (\ref{k-parts}) outside the integral go to zero  in the limits $\varphi_* \!\to\! -\infty$, $\varphi \!\to\! \infty$ because 
   $\hv$ does, while the  integral itself is bounded, so that $\left|\log\frac{\hI(\infty;\varepsilon)}{\hI(-\infty;\varepsilon)}\right|< N\varepsilon^k$ for some constant $N>0$. Arguing as above we conclude the proof of  (\ref{AdiabaticInv-k}).

\bigskip
{\bf Funding:} \
This research did not receive any specific grant from funding agencies in the public, commercial, or not-for-profit sectors.

\subsubsection*{Acknowledgments}

Work done also in the framework of the activities of G. Fiore within GNFM.

\end{document}